\documentclass[12pt]{iopart}
\usepackage{amsfonts}
\usepackage{amsbsy}

\eqnobysec

\begin{document}

\title{Explicit solutions for a (2+1)-dimensional Toda-like chain.}

\author{V.E. Vekslerchik}

\address{
Institute for Radiophysics and Electronics of NAS of Ukraine,
\\12, Proskura st., Kharkov, 61085, Ukraine}

\ead{vekslerchik@yahoo.com}

\begin{abstract}
We consider a (2+1)-dimensional Toda-like chain which can be viewed as a 
two-dimensional generalization of the Wu-Geng model and which is closely 
related to the two-dimensional Volterra, two-dimensional Toda and 
relativistic Toda lattices.
In the framework of the Hirota direct approach, we present equations 
describing this model as a system of bilinear equations that belongs to the 
Ablowitz-Ladik hierarchy. 
Using the Jacobi-like determinantal identities and the Fay identity for the 
theta-functions, we derive its Toeplitz, dark-soliton and quasiperiodic 
solutions as well as the similar set of solutions for the two-dimensional 
Volterra chain.
\end{abstract}



\submitto{\JPA}

\maketitle

\renewcommand{\vec}[1]{\boldsymbol{#1}}

\newcommand{\mytau}[1]{%
  \ifcase #1 \tau
  \or \omega
  \or \hat\tau
  \or \hat\omega
  \or \check\tau
  \or \check\omega
  \else ?\fi}

\newcommand{\mypretau}[1]{%
  \ifcase #1 T
  \or W
  \or ?
  \or \hat{W}
  \or \check{T}
  \else ?\fi}

\newcommand{\mydet}[3]{\mathsf{#1}^{#2}_{#3}}

\newcommand{\mymatrix}[1]{{\mathrm{#1}}}

\newcommand{\myomega}{\mathsf{\Omega}}

\newcommand{\myZero}[3]{%
  \ifcase #1 \mathbf{W}^{#2}_{#3}
  \or \mathbf{X}^{#2}_{#3}
  \or \mathbf{Y}^{#2}_{#3}
  \or \mathbf{Z}^{#2}_{#3}
  \or \mathbf{I}^{#2}_{#3}
  \or \mathbf{J}^{#2}_{#3}
  \or \mathbf{\bar{I}}^{#2}_{#3}
  \or \mathbf{\bar{J}}^{#2}_{#3}
  \else \stackrel{#1}{\mathfrak{O}}\!{}^{#2}_{#3} \fi}

\newcommand{\myzero}[3]{%
  \ifcase #1 \mathbf{w}^{#2}_{#3}
  \or \mathbf{x}^{#2}_{#3}
  \or \mathbf{y}^{#2}_{#3}
  \or \mathbf{z}^{#2}_{#3}
  \else \stackrel{#1}{\mathfrak{o}}\!{}^{#2}_{#3} \fi}

\newcommand{\mybarzero}[3]{%
  \ifcase #1 \bar\mathbf{w}^{#2}_{#3}
  \or \bar\mathbf{x}^{#2}_{#3}
  \or \bar\mathbf{y}^{#2}_{#3}
  \or \bar\mathbf{z}^{#2}_{#3}
  \else \stackrel{#1}{\bar\mathfrak{o}}\!{}^{#2}_{#3} \fi}

\section{Introduction.}

In this paper we consider a (2+1)-dimensional Toda-like chain
\begin{equation}
  \Delta u_{n} = 
  \left( \nabla u_{n}, \nabla u_{n} \right) 
  \left( 
    \frac{ 1 }{ u_{n} - u_{n+1} } 
    + 
    \frac{ 1 }{ u_{n} - u_{n-1} } 
  \right), 
\label{se-vect}
\end{equation}
where $u_{n}=u_{n}(x,y)$, 
$\Delta$ and $\nabla$ are the two-dimensional Laplacian and gradient. 
This equation appears in the literature in various contexts. 
For example, it is known to describe the B\"acklund transformations of a 
Heisenberg-like magnetics (see \cite{SY97}) and Laplace transformations of 
hydrodynamic-type systems in Riemann invariants \cite{F97}.
In a recent paper \cite{PV11} this system was shown to describe 
$O(3)$ $\sigma$-fields 
(three-component vectors $\vec{\sigma}_{n}$ of unit length) 
coupled by a nearest-neighbour Heisenberg-like interaction
such as, \textit{e.g.}, graphite-like magnetics when the spins 
inside one layer 
are governed by the Landau-Lifshitz theory with an effective Heisenberg 
interaction between adjacent layers: 
$\mathcal{H} = \mathcal{H}_{\mathrm{LL}} + \mathcal{H}_{\mathrm{H}} $
where 
$\mathcal{H}_{\mathrm{LL}} = \sum_{n}\mathcal{E}_{n}$
with
\begin{equation}
  \mathcal{E}_{n} = 
  \int\limits_{\mathbb{R}^{2}} dx \, dy \;
  \left( \nabla\vec{\sigma}_{n}, \nabla\vec{\sigma}_{n} \right)
\end{equation}
and 
$\mathcal{H}_{\mathrm{H}} = \frac{1}{2} \sum_{n}\sum_{p = n \pm 1} \mathcal{U}_{np}$
with 
\begin{equation}
  \mathcal{U}_{np}
  = 
  g^{2} 
  \int\limits_{\mathbb{R}^{2}} dx \, dy \;
  \ln\biggl( 1 + \left( \vec{\sigma}_{n}, \vec{\sigma}_{p} \right)\biggr). 
\end{equation}

A remarkable feature of the model \eref{se-vect}, which we will write  
in terms of the complex variables $z=x+iy$ and $\bar{z}=x-iy$,
\begin{equation}
  \partial\bar\partial u_{n} = 
  \left( \partial u_{n} \right) 
  \left( \bar\partial u_{n} \right) 
  \left( 
    \frac{ 1 }{ u_{n} - u_{n+1} } 
    + 
    \frac{ 1 }{ u_{n} - u_{n-1} } 
  \right) 
\label{se}
\end{equation}
where $\partial$ and $\bar\partial$ stand for 
$\partial / \partial z$ and $\partial / \partial \bar{z}$, is that it 
can be viewed as a connecting link between almost all Toda-like chains. 
In the following section, we present its relationships with 
the Wu-Geng chain (WGC) \cite{WG98},
the `two-dimensional Volterra equation' (2DVE) \cite{LSS80,LSS81}, 
the two-dimensional Toda lattice (2DTL) \cite{M79} 
and the relativistic Toda chain (RTC) \cite{R90,BR89a,RB89b,KMZ97} 

The main goal of this work is to obtain some explicit solutions for \eref{se}. 
We will not elaborate the inverse scattering transform or the 
algebro-geometric approach from scratch. Instead, we use the links between our 
equation and the 2DTL together with the results \cite{V95} that give us 
possibility of bilinearizing \eref{se} and reducing it in section 
\ref{sec-bilin} to a system that belongs 
to the Ablowitz-Ladik hierarchy (ALH) \cite{AL75}. 
Starting from the structure of the already known solutions for the ALH, we 
derive in sections \ref{sec-toeplitz} and \ref{sec-dark} Toeplitz and 
dark-soliton solutions directly from some determinantal identities. 
In section \ref{sec-qps}, we use the Fay identity for the $\theta$-functions 
to derive the quasiperiodic solutions. 
Finally, in section \ref{sec-tdve} the results of sections 
\ref{sec-toeplitz}--\ref{sec-qps} are used to 
obtain the Toeplitz, dark-soliton and periodic solutions for the 2DVE.

\section{WGC, 2DVE, 2DTL and RTC. \label{sec-etal}}

The main equations of this paper can be viewed as a straightforward 
generalization of the WGC to $(2+1)$ dimensions. Indeed, from equations (2) 
of \cite{WG98}, 
\begin{equation}
  u_{nt} = 
  \frac{ 1 }{ v_{n+1} }
  - 
  \frac{ 1 }{ v_{n} },
\qquad
v_{nt} = 
  \frac{ 1 }{ u_{n} }
  - 
  \frac{ 1 }{ u_{n-1} } 
\end{equation}
where $u_{nt}$ stands for $d u_{n} / dt$,
one can obtain that functions $w_{n}$ defined by 
$v_{n} = w_{n} - w_{n-1}$ satisfy 
\begin{equation}
  w_{nt} = \frac{ 1 }{ u_{n} } + C, 
  \qquad C=\mbox{constant}
\end{equation}
which leads, after setting $C=0$, to 
\begin{equation}
  w_{ntt} = 
  w_{nt}^{2} 
  \; 
  \frac{w_{n+1} - 2 w_{n} + w_{n-1} }
       {\left( w_{n+1} - w_{n} \right)\left( w_{n} - w_{n-1} \right)}.
\end{equation}
Clearly, this equation coincides with \eref{se} after replacing 
$\partial$ and $\bar\partial$ with $d/dt$.

On the other hand, rewriting equations \eref{se} in terms of the variables 
\begin{equation}
  a_{n} = \frac{ i\partial u_{n} }{ u_{n+1} - u_{n} }, 
  \qquad
  b_{n} = \frac{ -i\bar\partial u_{n} }{ u_{n} - u_{n-1} } 
\label{subst-ab-u}
\end{equation}
one arrives at the system 
\begin{equation}
  \left\{ 
  \begin{array}{lcl}
  i \bar\partial a_{n} & = & a_{n} \left( b_{n+1} - b_{n} \right) 
  \\[2mm]
  i \partial b_{n} & = & b_{n} \left( a_{n-1} - a_{n} \right) 
  \end{array} 
  \right.
\label{se-ab}
\end{equation}
which is closely related to the 2DTL: 
the quantities $f_{n}$ defined by 
\begin{equation}
  f_{n} = \ln a_{n}b_{n}
\end{equation}
satisfy
\begin{equation}
  \partial\bar\partial f_{n} = e^{f_{n+1}} - 2 e^{f_{n}} + e^{f_{n-1}}. 
\end{equation}
System \eref{se-ab} is known since the works of Leznov, Savel'ev and Smirnov 
\cite{LSS80,LSS81} where the authors demonstrated that this system, 
which was named the `two-dimensional Volterra equation', 
represents the B\"acklund transformations for the 2DTL and constructed its general 
solutions in the finite case using the results of \cite{LS79,L80}.

Another model that we would like to discuss is the RTC  
\cite{R90,BR89a,RB89b,KMZ97} which can be presented as a Hamiltonian system 
\begin{equation}
  i \partial q_{n} = \partial \mathcal{H} / \partial p_{n}, 
  \qquad
  i \partial p_{n} = - \partial \mathcal{H} / \partial q_{n}  
\label{rtc-ham}
\end{equation}
with 
\begin{equation}
  \mathcal{H} 
  = 
  \sum_{n} e^{ p_{n} } \left( e^{ q_{n+1}-q_{n} } - 1 \right).  
\end{equation}
Equations \eref{rtc-ham}
\begin{equation}
  \left\{ 
  \begin{array}{lcl}
  i \partial q_{n} 
  & = & 
  e^{ p_{n} } \left( e^{ q_{n+1} - q_{n} } - 1 \right) 
  \\[2mm]  
  i \partial p_{n} 
  & = & 
  e^{ p_{n} + q_{n+1} - q_{n} } 
  - 
  e^{ p_{n-1} + q_{n} - q_{n-1} } 
  \end{array} 
  \right.
\label{rtc-qp-pos}
\end{equation}
rewritten in terms of the variables 
\begin{equation}
  a_{n} = e^{ p_{n} + q_{n+1} - q_{n} }, 
  \qquad
  b_{n} = e^{ -p_{n} } 
\label{subst-ab-qp}
\end{equation}
are
\begin{equation}
  \left\{ 
  \begin{array}{lcl}
  i\partial \ln a_{n} & = & 
  a_{n+1} - a_{n-1} - 1/b_{n+1} + 1/b_{n} 
  \\ 
  i\partial \ln b_{n} & = & 
  a_{n-1} - a_{n}. 
  \end{array} 
  \right.
\label{rtc-ab-pos}
\end{equation}
One can easily note that the second of the above equations coincides with the 
second equation of \eref{se-ab}. In a similar manner, starting from the 
Hamiltonian system 
\begin{equation}
  i \bar\partial q_{n} = \partial \bar{\mathcal{H}} / \partial p_{n}, 
  \qquad
  i \bar\partial p_{n} = - \partial \bar{\mathcal{H}} / \partial q_{n} 
\end{equation}
with 
\begin{equation}
  \bar{\mathcal{H}}  
  = 
  \sum_{n} e^{ -p_{n} } \left( e^{ q_{n}-q_{n-1} } - 1 \right)  
\end{equation}
that has the form 
\begin{equation}
  \left\{ 
  \begin{array}{lcl}
  i \bar\partial q_{n} 
  & = & 
  e^{ -p_{n} } \left( 1 - e^{ q_{n} - q_{n-1} } \right) 
  \\[2mm]  
  i \bar\partial p_{n} 
  & = & 
  e^{ -p_{n+1} + q_{n+1} - q_{n} } 
  - 
  e^{ -p_{n} + q_{n} - q_{n-1} } 
  \end{array} 
  \right.
\label{rtc-qp-neg}
\end{equation}
one arrives, after using substituting \eref{subst-ab-qp}, at the system
\begin{equation}
  \left\{ 
  \begin{array}{lcl}
  i\bar\partial \ln a_{n} & = & 
  b_{n+1} - b_{n} 
  \\ 
  i\partial \ln b_{n} & = & 
  a_{n-1}b_{n-1}b_{n} - a_{n}b_{n}b_{n+1} 
  \end{array} 
  \right.
\label{rtc-ab-neg}
\end{equation}
whose first equation is the first one from \eref{se-ab}.
The relationship between the two RTCs, \eref{rtc-ab-pos} and \eref{rtc-ab-neg}, 
and our system \eref{se-ab} becomes transparent if one considers RTCs from the 
zero-curvature viewpoint based on the scattering problem \cite{BR89a,RB89b,KMZ97}
\begin{equation}
  a_{n}\psi_{n+1} - b_{n}^{-1} \psi_{n} 
  = \Lambda \left( \psi_{n} - \psi_{n-1} \right).
\label{rtc-sp}
\end{equation}
System \eref{rtc-ab-pos} is the compatibility condition of \eref{rtc-sp} and 
\begin{equation}
  i \partial \psi_{n} = 
  a_{n} \left( \psi_{n+1} - \psi_{n} \right) 
\label{rtc-evol-pos}
\end{equation}
while \eref{rtc-ab-neg} plays the same role for \eref{rtc-sp} and 
\begin{equation}
  i \bar\partial \psi_{n} = 
  b_{n} \left( \psi_{n-1} - \psi_{n} \right). 
\label{rtc-evol-neg}
\end{equation}
As to our system, \eref{se-ab}, it ensures the consistency of 
\eref{rtc-evol-pos} taken together with \eref{rtc-evol-neg}, 
\begin{equation}
  \left\{ 
  \begin{array}{lcl}
  i \partial \psi_{n} & = & 
  a_{n} \left( \psi_{n+1} - \psi_{n} \right) 
  \\ 
  i \bar\partial \psi_{n} & = & 
  b_{n} \left( \psi_{n-1} - \psi_{n} \right). 
  \end{array} 
  \right.
\label{rtc-evol-pn}
\end{equation}
To summarize, equations that are the subject of this paper can be viewed as 
describing the commutativity of two relativistic Toda flows. 
It should be noted that this commutativity leads, again, to the 2DTL. Indeed, 
considering equations \eref{rtc-qp-pos} combined with \eref{rtc-qp-neg} one 
can obtain by direct calculations that functions $q_{n}$ satisfy 
\begin{equation}
  \partial\bar\partial q_{n} 
  = 
  \exp\left( q_{n+1} - q_{n} \right) 
  - 
  \exp\left( q_{n} - q_{n-1} \right) 
\end{equation}
which is another form of the 2DTL.

Finally, comparing \eref{subst-ab-u} with \eref{rtc-evol-pn} one can conclude 
that solutions for our system, $u_{n}$, are nothing but solutions for the 
auxiliary linear problems for the 2DVE; in other words, the $u_{n}$-model is 
dual to $\left(a_{n},b_{n}\right)$-model.

\section{Bilinearization. \label{sec-bilin}}

To bilinearize our equation, that can be written as
\begin{equation}
  \left\{
  \begin{array}{rcl} 
  \partial\bar\partial u_{n} & = & p_{n} \left( u_{n+1} - 2 u_{n} + u_{n-1} \right) 
  \\[2mm] 
  \left( \partial u_{n} \right)\left( \bar\partial u_{n} \right) & = &
  p_{n} \left( u_{n+1} - u_{n} \right) \left( u_{n} - u_{n-1} \right) 
  \end{array}
  \right.
\label{toda-u}
\end{equation}
we start with the 2DTL, assuming 
\begin{equation}
  p_{n} = \frac{ \tau_{n-1}\tau_{n+1} }{ \tau_{n}^{2} }
\end{equation}
where $\tau_{n}$ is a solution for 
\begin{equation}
  \partial \bar\partial \ln\tau_{n} 
  = 
  \frac{ \tau_{n-1}\tau_{n+1} }{ \tau_{n}^{2} }
\end{equation}
or, in the bilinear form, 
\begin{equation}
  D \bar{D} \, \tau_{n} \cdot \tau_{n} = 2\tau_{n-1}\tau_{n+1} 
\end{equation}
where $D$ and $\bar{D}$ are the Hirota operators corresponding to 
$\partial$ and $\bar\partial$: 
$D \, a \cdot b =  (\partial a)b - a(\partial b)$,
$\bar{D} \, a \cdot b =  (\bar\partial a)b - a(\bar\partial b)$.
 
One can easily obtain a wide range of solutions
for the first equation of \eref{toda-u} by differentiating the 2DTL 
tau-function $\tau_{n}$ with respect to \emph{any} parameter, 
\begin{equation}
  u_{n} = \frac{ \partial }{ \partial \zeta} \ln\tau_{n}. 
\end{equation}
Since $\tau_{n}$ can be considered as a solution for \emph{all} equations of 
the 2DTL hierarchy, i.e. as a function of an infinite number of `times', 
$\tau_{n} = 
 \tau_{n}\left( t_{1}, t_{2}, ... \bar{t}_{1}, \bar{t}_{2}, ... \right)$, 
$t_{1}=z$, $\bar{t}_{1}=\bar{z}$,
the last formula can be generalized as follows:
\begin{equation}
  u_{n} = 
  \sum_{j=1}^{\infty} \left(
    c_{j} \frac{ \partial \ln\tau_{n} }{ \partial t_{j}}  
    + 
    \bar{c}_{j} \frac{ \partial \ln\tau_{n} }{ \partial \bar{t}_{j}}  
  \right) 
\end{equation}
with arbitrary constants $c_{j}$ and $\bar{c}_{j}$. 
However, this apparently most straightforward approach is rather hard to 
implement because the second equation of \eref{toda-u}, 
which plays the role of the restriction for the set 
$c_{j}$ and $\bar{c}_{j}$, is a bilinear system which we cannot solve. 
That is why we will use another way to 
deal with our equations. 
The key point, that will be proved below, is that 
$u_{n}$ can be composed of \emph{two} solutions for the 2DTL equation:
\begin{equation}
  u_{n} = \frac{ \mytau1_{n} }{ \mytau0_{n} } 
\end{equation}
where
\numparts 
\begin{eqnarray}
  {\scriptstyle \frac{1}{2}} \, D \bar{D} \, \mytau1_{n} \cdot \mytau1_{n} 
  & = & 
  \mytau1_{n-1} \mytau1_{n+1} 
  + 
  \lambda \, \mytau1_{n}^{2} 
  \\
  {\scriptstyle \frac{1}{2}} \, D \bar{D} \, \mytau0_{n} \cdot \mytau0_{n} 
  & = & 
  \mytau0_{n-1} \mytau0_{n+1} 
  + 
  \lambda \, \mytau0_{n}^{2} 
  \end{eqnarray}
\endnumparts 
and $\lambda$ is a unimportant constant that can be eliminated by 
multiplying $\mytau1_{n}$ and $\mytau0_{n}$ by $\exp(-\lambda z \bar{z})$ 
without modifying $u_{n}$.
Calculating the first derivatives of $u_{n}$, 
\begin{equation}
  \partial u_{n} = 
  \frac{ 1 }{ \mytau0_{n}^{2} } D \, \mytau1_{n} \cdot \mytau0_{n}, 
\qquad
  \bar\partial u_{n} = 
  \frac{ 1 }{ \mytau0_{n}^{2} } \bar{D} \, \mytau1_{n} \cdot \mytau0_{n} 
\end{equation} 
and its Laplacian, 
\begin{equation}
  \partial \bar\partial u_{n} = 
  \frac{ 1 }{ \mytau0_{n}^{2} } D \bar{D} \, \mytau1_{n} \cdot \mytau0_{n} 
  - 
  \frac{ u_{n} }{ \mytau0_{n}^{2} } \; D \bar{D} \, \mytau0_{n} \cdot \mytau0_{n} 
\end{equation}
one can rewrite \eref{toda-u} as 
\begin{equation}
  \left\{
  \begin{array}{rcl} 
  \left( D \bar{D} - 2\lambda \right) \mytau1_{n} \cdot \mytau0_{n} 
  & = & 
  \mytau0_{n-1} \mytau1_{n+1} + \mytau0_{n+1} \mytau1_{n-1} 
  \\
  \left( D \, \mytau1_{n} \cdot \mytau0_{n} \right) 
  \left( \bar{D} \, \mytau1_{n} \cdot \mytau0_{n} \right) 
  & = &
  \left( \mytau0_{n-1}\mytau1_{n} - \mytau0_{n}\mytau1_{n-1} \right) 
  \left( \mytau0_{n}\mytau1_{n+1} - \mytau0_{n+1}\mytau1_{n} \right). 
  \end{array}
  \right.
\label{syst-a}
\end{equation}
The next step is to split the second equation of the above 
system in two bilinear equations. Introducing new $\tau$-functions 
$\mytau2_{n}$, $\mytau3_{n}$, $\mytau4_{n}$ and $\mytau5_{n}$ 
by
\numparts
\begin{eqnarray}
  \phantom{-}iD \, \mytau1_{n} \cdot \mytau0_{n} & = & 
  \mytau2_{n} \mytau3_{n} 
  \\
  -i\bar{D} \, \mytau1_{n} \cdot \mytau0_{n} & = & 
  \mytau4_{n} \mytau5_{n} 
\end{eqnarray}
\endnumparts
and noting that the r.h.s. the second equation of \eref{syst-a} can be presented as 
$X_{n} X_{n-\delta}$
where 
\begin{equation}
  X_{n} = \mytau0_{n}\mytau1_{n+\delta} - \mytau0_{n+\delta}\mytau1_{n}, 
  \qquad 
  \delta = \pm 1 
\label{def-delta}
\end{equation}
it is possible to achieve our goal by setting
\begin{equation}
  \mytau5_{n} = \mytau3_{n-\delta}, 
  \qquad
  \mytau4_{n} = \mytau2_{n+\delta} 
\end{equation}
which leads to 
\begin{equation}
  \left\{
  \begin{array}{rcl} 
  \mytau0_{n}\mytau1_{n+\delta} - \mytau0_{n+\delta}\mytau1_{n} 
  & = & 
  \mytau4_{n} \, \mytau3_{n} 
  \\[2mm]
  iD \, \mytau1_{n} \cdot \mytau0_{n} & = & 
  \mytau4_{n-\delta} \, \mytau3_{n} 
  \\[1mm]
  -i\bar{D} \, \mytau1_{n} \cdot \mytau0_{n} & = & 
  \mytau4_{n} \, \mytau3_{n-\delta}. 
  \end{array} 
  \right.
\end{equation}
Now we have to close this system adding the equations describing the 
dependence of the new $\tau$-functions on the variables $z$ and 
$\bar{z}$. These equations should be (1) bilinear, (2) compatible and (3) 
imply as a consequence the first equation of \eref{syst-a}. 
The resulting system can be written as the union of the `positive' part,
\begin{equation}
  \left\{
  \begin{array}{lcl}
  0 & = & 
  iD \, \mytau1_{n} \cdot \mytau0_{n} 
  - \mytau4_{n-\delta} \mytau3_{n} 
  \\[1mm]
  0 & = & 
  iD \, \mytau4_{n} \cdot \mytau0_{n} 
  + \mytau4_{n-\delta} \mytau0_{n+\delta} 
  \\[1mm]
  0 & = & 
  iD \, \mytau1_{n} \cdot \mytau3_{n-\delta} 
  + \mytau1_{n-\delta} \mytau3_{n}
  , 
  \end{array}
  \right.
\label{main-pos}
\end{equation}
the `negative' one, 
\begin{equation}
  \left\{ 
  \begin{array}{lcl}
  0 & = & 
  i \bar{D} \, \mytau1_{n} \cdot \mytau0_{n} 
  + \mytau4_{n} \mytau3_{n-\delta} 
  \\[1mm]
  0 & = & 
  i \bar{D} \, \mytau4_{n-\delta} \cdot \mytau0_{n} 
  + \mytau4_{n} \mytau0_{n-\delta} 
  \\[1mm]
  0 & = & 
  i \bar{D} \, \mytau1_{n} \cdot \mytau3_{n} 
  + \mytau1_{n+\delta} \mytau3_{n-\delta}  
  , 
  \end{array} 
  \right.
\label{main-neg}
\end{equation}
combined with the restriction
\begin{equation}
  0 =  
    \mytau0_{n} \mytau1_{n+\delta} 
  - \mytau0_{n+\delta} \mytau1_{n} 
  - \mytau4_{n} \mytau3_{n}. 
\label{main-alh}
\end{equation}
One can verify that these bilinear equations indeed lead to the solution of 
our problem. As to their compatibility (that can be checked directly, but 
after rather long and tedious calculations), we would like to note that they 
are a part of the extended ALH \cite{V02}. Actually, we have passed from 
$\tau$-functions of the 2DTL to ones of the ALH using the results of \cite{V95}. 
In other words, the above proceeding can be viewed as an explanation of the 
following fact, that can be verified by straightforward algebra: 
the compatible system \eref{main-pos}--\eref{main-alh} which belongs to the 
extended ALH provides solutions for the system \eref{toda-u}. The reference 
to the ALH not only resolves the question of compatibility, but also gives 
us possibility of using the already known solutions for the Ablowitz-Ladik 
equations to build ones for the problem we are dealing with. 

Definitions \eref{subst-ab-u} and the above equations yield the following 
expressions for solutions of the 2DVE: 
\numparts
\begin{eqnarray}
  \delta = 1: 
  & \qquad &
  a_{n} = \frac{ \mytau4_{n-1} \mytau0_{n+1} }{ \mytau4_{n} \mytau0_{n} }, 
  \quad
  b_{n} = \frac{ \mytau0_{n-1} \mytau4_{n} }{ \mytau4_{n-1} \mytau0_{n} } 
\label{ab-tau-a}
\\
  \delta = -1: 
  & \qquad &
  a_{n} = - \frac{ \mytau3_{n} \mytau0_{n+1} }{ \mytau0_{n} \mytau3_{n+1} }, 
  \quad
  b_{n} = - \frac{ \mytau0_{n-1} \mytau3_{n+1} }{ \mytau0_{n} \mytau3_{n} } 
\label{ab-tau-b}
\end{eqnarray}
\endnumparts
that will be used in section \ref{sec-tdve}.

\section{Toeplitz solutions. \label{sec-toeplitz}} 

This type of solutions can be constructed of the Toeplitz determinants
\begin{equation}
  \mydet{A}{k}{\ell} = 
  \det\left| \alpha_{k+a-b} \right|_{a,b=1,...,\ell}.
\label{det-A}
\end{equation}
The main idea is that among various determinantal identities one can find ones 
that are similar to the equations we want to solve. 
In the appendix we derive the necessary formulae by applying the Jacobi identity 
to the determinants \eref{det-A} and the framed ones, 
\begin{equation}
  \mydet{F}{k}{\ell+1}(\zeta) = 
  \left|
  \begin{array}{ccccc} 
    1 & \zeta & \zeta^{2} & \dots & \zeta^{\ell} \cr
    \alpha_{k-1} & \alpha_{k} & \alpha_{k+1} & \dots & \alpha_{k + \ell -1} \cr
    \vdots & \vdots & \vdots & \ddots & \vdots
  \end{array} 
  \right|
\end{equation}
(see \eref{app-pos-a}--\eref{app-neg-c}).
These formulae can then be rewritten in terms of the 
$\ell$-order determinants $\mydet{B}{k}{\ell}$, similar to \eref{det-A}, 
instead of the $(\ell+1)$-order ones, $\mydet{F}{k}{\ell+1}(\zeta)$, 
\begin{equation}
  \mydet{F}{k}{\ell+1}(\zeta) 
  = 
  (-\zeta)^{\ell} \mydet{B}{k-1}{\ell}(\zeta), 
\label{def-B}
\end{equation}
where 
\begin{equation}
  \mydet{B}{k}{\ell}(\zeta) = 
  \det\left| \beta_{k+a-b}(\zeta) \right|_{a,b=1,...,\ell} 
\end{equation}
with 
\begin{equation}
  \beta_{k}(\zeta) = \alpha_{k} - \zeta^{-1} \alpha_{k+1}. 
\end{equation}
The set of identities that we need to solve our equations is 
\numparts 
\begin{eqnarray}
\label{AB-pos-a}
  0 & = & 
  D_{+} \, \mydet{B}{k}{\ell} \cdot \mydet{A}{k}{\ell} 
  - \zeta^{-1} \mydet{A}{k+1}{\ell+1} \, \mydet{B}{k-1}{\ell-1} 
\\[2mm]
  0 & = & 
  D_{+} \, \mydet{B}{k}{\ell} \cdot \mydet{A}{k}{\ell+1} 
  + \mydet{A}{k+1}{\ell+1} \, \mydet{B}{k-1}{\ell} 
\\[2mm]
  0 & = & 
  D_{+} \, \mydet{B}{k}{\ell} \cdot \mydet{A}{k+1}{\ell} 
  - \mydet{A}{k+1}{\ell+1} \, \mydet{B}{k}{\ell-1} 
\label{AB-pos-c}
\end{eqnarray}
\endnumparts 
and
\numparts 
\begin{eqnarray}
\label{AB-neg-a}
  0 & = & 
  D_{-} \, \mydet{B}{k}{\ell} \cdot \mydet{A}{k}{\ell} 
  + \zeta^{-1} \mydet{A}{k}{\ell+1} \, \mydet{B}{k}{\ell-1} 
\\[2mm]
  0 & = & 
  D_{-} \, \mydet{B}{k}{\ell} \cdot \mydet{A}{k+1}{\ell} 
  + \mydet{A}{k}{\ell+1} \, \mydet{B}{k+1}{\ell-1} 
\\[2mm]
  0 & = & 
  D_{-} \, \mydet{B}{k}{\ell} \cdot \mydet{A}{k+1}{\ell+1} 
  + \mydet{A}{k}{\ell+1} \, \mydet{B}{k+1}{\ell} 
\label{AB-neg-c}
\end{eqnarray}
\endnumparts 
together with 
\begin{equation}
  0 =  
  \mydet{A}{k+1}{\ell} \mydet{B}{k}{\ell}  
  - \mydet{A}{k}{\ell} \mydet{B}{k+1}{\ell} 
  + \zeta^{-1} \mydet{A}{k+1}{\ell+1} \mydet{B}{k}{\ell-1}. 
\label{AB-alg-c}
\end{equation}
Here, $D_{\pm}$ stand for the Hirota operators corresponding to 
$\partial_{\pm}$ and 
$\partial_{\pm}\alpha_{k} = \alpha_{k \pm 1}$.

Now, the problem of finding solutions for \eref{main-pos}--\eref{main-alh} 
becomes a `combinatorial' one: one has to select 
$\mytau0_{n}$, $\mytau1_{n}$, $\mytau4_{n}$, $\mytau3_{n}$, from 
$\mydet{A}{k}{\ell}$, $\mydet{B}{k}{\ell}$ with different $k$ and $\ell$. 
Below, we obtain three families of solutions: infinite (with respect to $n$), 
semi-infinite and finite ones.

\subsection{Infinite chain.} 

This type of solutions corresponds to the following choice of the 
$\tau$-functions: 
\begin{equation}
  \mytau0_{n} = \mytau0_{*} \, \mydet{A}{n}{\ell},
  \qquad 
  \mytau1_{n} = \mytau1_{*} \, \mydet{B}{n}{\ell}
\label{toep-inf-a}
\end{equation}
and 
\begin{equation}
  \mytau4_{n} = \mytau4_{*} \, \mydet{B}{n}{\ell-1}, 
  \qquad 
  \mytau3_{n} = \mytau3_{*} \, \mydet{A}{n+1}{\ell+1} 
\label{toep-inf-c}
\end{equation}
with constant $\mytau0_{*}$, $\mytau1_{*}$, $\mytau4_{*}$, $\mytau3_{*}$. 
From \eref{AB-alg-c}, one can immediately derive that 
\begin{equation}
  0 = 
  \mytau0_{n} \mytau1_{n+1} 
  - \mytau0_{n+1} \mytau1_{n} 
  - \mytau4_{n} \mytau3_{n} 
\end{equation}
provided 
\begin{equation} 
  \mytau0_{*} \mytau1_{*} 
  = 
  \zeta
  \mytau4_{*} \mytau3_{*}. 
\end{equation}
Thus, our $\tau$-functions solve \eref{main-alh} with $\delta=1$.
Furthermore, equations \eref{AB-pos-a}--\eref{AB-neg-c} lead to 
\numparts 
\begin{eqnarray}
  0 & = & 
  D_{+} \mytau1_{n} \cdot \mytau0_{n} 
  - 
  \mytau4_{n-1} \, \mytau3_{n} 
  \\[2mm] 
  0 & = & 
  D_{+} \mytau4_{n} \cdot \mytau0_{n} 
  + 
  \mytau4_{n-1} \, \mytau0_{n+1} 
  \\[2mm] 
  0 & = & 
  D_{+} \mytau1_{n} \cdot \mytau3_{n-1} 
  + 
  \mytau1_{n-1} \, \mytau3_{n} 
\end{eqnarray}
\endnumparts 
and 
\numparts 
\begin{eqnarray}
  0 & = & 
  D_{-} \, \mytau1_{n} \cdot \mytau0_{n} 
  + 
  \mytau3_{n-1} \, \mytau4_{n} 
  \\[2mm] 
  0 & = & 
  D_{-} \, \mytau4_{n-1} \cdot \mytau0_{n} 
  + 
  \mytau0_{n-1} \, \mytau4_{n}
  \\[2mm] 
  0 & = & 
  D_{-} \, \mytau1_{n} \cdot \mytau3_{n} 
  + 
  \mytau3_{n-1} \, \mytau1_{n+1}. 
\end{eqnarray}
\endnumparts 
It is clear that to complete solution for \eref{main-pos}--\eref{main-neg} 
one has to meet 
\begin{equation}
  \partial_{+} = i\partial,
  \qquad
  \partial_{-} = i\bar\partial 
\end{equation}
or, to take $\alpha_{k}$ to be solutions for the \textit{linear} system 
\begin{equation}
  \left\{ 
  \begin{array}{lcl} 
    i\partial\alpha_{k} & = & \alpha_{k+1} \\
    i\bar\partial\alpha_{k} & = & \alpha_{k-1}. 
  \end{array} 
  \right.
\label{alpha-infinite}
\end{equation}
The `symmetric' set of solutions can be obtained by using, instead of 
\eref{def-B}, another representation of the determinants 
$\mydet{F}{k}{\ell+1}(\zeta)$:
\begin{equation}
  \mydet{F}{k}{\ell+1}(\zeta) 
  = 
  \mydet{C}{k}{\ell}(\zeta) 
\label{def-C}
\end{equation}
where 
\begin{equation}
  \mydet{C}{k}{\ell}(\zeta) = 
  \det\left| \gamma_{k+a-b}(\zeta) \right|_{a,b=1,...,\ell} 
\end{equation} 
with
\begin{equation}
  \gamma_{k}(\zeta) = \alpha_{k} - \zeta \alpha_{k-1}. 
\end{equation}
The calculations similar to the ones presented above demonstrate that 
$\tau$-functions defined by 
\begin{equation}
  \mytau0_{n} =  
  \mytau0_{*} \, \mydet{A}{n}{\ell},
  \qquad
  \mytau1_{n} = 
  \mytau1_{*} \, \mydet{C}{n}{\ell}
\end{equation}
and 
\begin{equation}
  \mytau4_{n} = 
  \mytau4_{*} \, \mydet{C}{n+1}{\ell-1}, 
  \qquad 
  \mytau3_{n} = 
  \mytau3_{*} \, \mydet{A}{n}{\ell+1} 
\end{equation}
with 
\begin{equation}
  \zeta\mytau0_{*} \mytau1_{*} 
  = 
  - \mytau4_{*} \mytau3_{*} 
\end{equation}
solve \eref{main-pos}--\eref{main-alh} with $\delta=1$. 

Both these sets of solutions can be written as 
\begin{equation}
  u_{n}\left(z,\bar{z}\right) = 
  u_{*} \,
  \frac{ 
    \det\left| \alpha_{n+a-b}^{\pm}\left(z,\bar{z}\right) \right|_{a,b=1,...,\ell} 
  }{  
    \det\left| \alpha_{n+a-b}\left(z,\bar{z}\right) \right|_{a,b=1,...,\ell} 
  }
\label{u-toeplitz-inf}
\end{equation}
where $\ell$ is an arbitrary positive integer and 
the elements of the determinants are given by 
\numparts
\begin{eqnarray}
  \alpha_{k}\left(z,\bar{z}\right) 
  & = & 
  \int\nolimits_{\Gamma} dh \, \hat\alpha(h) \; 
  h^{k} \exp\left[ - i \Theta_{h}\left(z,\bar{z}\right) \right]
\label{sol-alpha}
\\
  \alpha_{k}^{\pm}\left(z,\bar{z}\right) 
  & = & 
  \int\nolimits_{\Gamma} dh \, \hat\alpha(h) 
  \left[ 1 - \left(h/\zeta\right)^{\pm 1} \right] 
  h^{k} \exp\left[ - i \Theta_{h}\left(z,\bar{z}\right) \right]
\label{sol-alpha-pm}
\end{eqnarray}
\endnumparts
(we replaced $\beta_{k}$ and $\gamma_{k}$ with $\alpha_{k}^{\pm}$) 
with arbitrary contour $\Gamma$, function $\hat\alpha(h)$ and  
constant $u_{*}$. The `dispersion law' $\Theta_{h}\left(z,\bar{z}\right)$ is 
given by 
\begin{equation}
  \Theta_{h}\left(z,\bar{z}\right) =  hz +  h^{-1}\bar{z}. 
\end{equation}  

As an example, let us consider one of the simplest solutions of 
\eref{alpha-infinite}. Noting that system \eref{alpha-infinite} leads to the 
Helmholtz equation, 
$\partial\bar\partial\alpha_{k} + \alpha_{k} = 0$,
and rewriting the latter using the polar coordinates, $z=re^{i\theta}$, one 
can obtain in a standard way 
\begin{equation}
  \alpha_{k} = \exp\left[ - i (\theta + \pi/2) k \right] J_{k}(2r) 
\end{equation}  
where $J_{k}$ is the $k$th Bessel function.
These solutions correspond to \eref{sol-alpha} and \eref{sol-alpha-pm}
with $\hat\alpha(h) = ( 2\pi i h )^{-1}$ and $\Gamma$ being the unit 
circumference: $\Gamma = \{ h : |h|=1 \}$. 
In the `elementary' case of $\ell=1$ expression \eref{u-toeplitz-inf} can be 
written as 
\begin{equation}
  u_{n} = 
  u_{*} 
  + u_{\scriptscriptstyle \pm} 
  \frac{ J_{n \pm 1}(2r) }{ J_{n}(2r) } 
  e^{\mp i\theta} 
\end{equation}
where $u_{*}$ and $u_{\scriptscriptstyle \pm}$ 
($u_{\scriptscriptstyle \pm} = \pm i u_{*}\zeta^{\mp 1}$) 
are arbitrary constants.
It is easy to see that solutions we have obtained are complex and singular, 
which, however, does not mean that they are non-physical. Say in the case of 
\cite{PV11}, that was discussed in the introduction, the spin components, 
$\vec{\sigma}_{n} = ( \sigma_{n}^{(1)}, \sigma_{n}^{(2)}, \sigma_{n}^{(3)} )$, 
are related to $u_{n}$ by 
$\sigma_{n}^{(1)} + i\sigma_{n}^{(2)} = 2u_{n} / (1 + |u_{n}|^{2} )$ and 
$\sigma_{n}^{(3)} = (1 - |u_{n}|^{2} ) / (1 + |u_{n}|^{2} )$. 
Thus, in the general case ($ \sigma_{n}^{(2)} \ne 0 $) $u_{n}$ is complex and 
singularities of $u_{n}$ correspond to the vertical (southward) orientation of 
$\vec{\sigma}_{n}$: 
$u_{n}=\infty \Leftrightarrow \sigma_{n}^{(1)} = \sigma_{n}^{(2)} = 0$ and 
$\sigma_{n}^{(3)} = -1$.

\subsection{Semi-infinite chain.} 

This type of solutions appears if one identify the index $n$ with the size of 
the determinants $\mydet{A}{k}{\ell}$, $\mydet{B}{k}{\ell}$.
Consider the functions 
$\mypretau0_{n}$, $\mypretau1_{n}$, $\mypretau4_{n}$, $\mypretau3_{n}$ 
given by 
\begin{equation}
  \mypretau0_{n} = 
  \mypretau0_{*} \, \mydet{A}{k}{n}, 
  \qquad 
  \mypretau1_{n} = 
  \mypretau1_{*} \, \mydet{B}{k}{n}
\end{equation}
and 
\begin{equation}
  \mypretau4_{n} = 
  \mypretau4_{*} \, \mydet{B}{k-1}{n}, 
  \qquad
  \mypretau3_{n} = 
  \mypretau3_{*} \, \mydet{A}{k+1}{n+1} 
\end{equation}
with constants 
$\mypretau0_{*}$, $\mypretau1_{*}$, $\mypretau4_{*}$, $\mypretau3_{*}$ 
being related by 
\begin{equation} 
  \mypretau0_{*} \mypretau1_{*}
  = 
  - \zeta \mypretau4_{*} \mypretau3_{*}. 
\end{equation}
It is straightforward to verify that they solve 
\begin{equation}
  0 = 
  \mypretau0_{n} \mypretau1_{n+1} 
  - \mypretau0_{n+1} \mypretau1_{n} 
  - \mypretau4_{n} \mypretau3_{n} 
\end{equation}
as well as 
\begin{equation}
  \begin{array}{lcl}
  0 & = & 
  D_{+} \mypretau1_{n} \cdot \mypretau0_{n} 
  + \mypretau4_{n-1} \, \mypretau3_{n} 
  \\[2mm] 
  0 & = & 
  D_{+} \mypretau4_{n} \cdot \mypretau0_{n} 
  - \mypretau4_{n-1} \, \mypretau0_{n+1} 
  \\[2mm] 
  0 & = & 
  D_{+} \mypretau1_{n} \cdot \mypretau3_{n-1} 
  - \mypretau1_{n-1} \, \mypretau3_{n} 
  \end{array} 
\end{equation}
and 
\begin{equation}
  \begin{array}{lcl}
  0 & = & 
  \left( D_{-} + \zeta^{-1} \right) \mypretau1_{n} \cdot \mypretau0_{n} 
  + 
  \mypretau3_{n-1} \, \mypretau4_{n} 
  \\[2mm] 
  0 & = & 
  \left( D_{-} + \zeta^{-1} \right) \mypretau4_{n-1} \cdot \mypretau0_{n} 
  + 
  \mypretau0_{n-1} \, \mypretau4_{n} 
  \\[2mm] 
  0 & = & 
  \left( D_{-} + \zeta^{-1} \right) \mypretau1_{n} \cdot \mypretau3_{n} 
  + 
  \mypretau3_{n-1} \, \mypretau1_{n+1} 
  \end{array}
\end{equation}
that almost coincide with \eref{main-pos}--\eref{main-alh} 
for $\delta=1$ after identifying 
\begin{equation}
  \partial = i \partial_{+}, 
  \qquad
  \bar\partial = - i \partial_{-}. 
\label{alpha-semi}
\end{equation}
The extra terms can be eliminated by $\exp(i\bar{z}/\zeta)$ factor and one 
arrives at the following solutions: 
\begin{equation}
\label{toep-semi-a}
  \mytau0_{n} = 
  \mytau0_{*} \, \exp(i\bar{z}/\zeta) \, \mydet{A}{k}{n}, 
  \qquad 
  \mytau1_{n} =  
  \mytau1_{*} \, \mydet{B}{k}{n}
\end{equation}
and 
\begin{equation}
\label{toep-semi-c}
  \mytau4_{n} = 
  \mytau4_{*} \, \mydet{B}{k-1}{n}, 
  \qquad 
  \mytau3_{n} =  
  \mytau3_{*} \, \exp(i\bar{z}/\zeta) \, \mydet{A}{k+1}{n+1} 
\end{equation}
where $\mytau0_{*}$, $\mytau1_{*}$, $\mytau4_{*}$, $\mytau3_{*}$ are arbitrary 
constants and 
\begin{equation}
  \zeta 
  = 
  - \frac{  \mytau0_{*} \mytau1_{*} }{ \mytau4_{*} \mytau3_{*} }. 
\end{equation}

In a similar way one can derive the `symmetric' set of solutions using the 
determinants $\mydet{C}{k}{n}$. 
These two sets of the Toeplitz solutions $u_{n}$ can be written as
\begin{equation}
  u_{n}\left(z,\bar{z}\right) = 
  u_{*}
  e^{ i\phi_{\pm}\left(z,\bar{z}\right) } \;
  \frac{ 
    \det\left| \alpha_{k+a-b}^{\pm}\left(z,\bar{z}\right) \right|_{a,b=1,...,n} 
  }{  
    \det\left| \alpha_{k+a-b}\left(z,\bar{z}\right) \right|_{a,b=1,...,n} 
  }
\label{u-sol-semi}
\end{equation}
with an arbitrary positive integer $k$.
The elements of the determinants are given again by \eref{sol-alpha} and 
\eref{sol-alpha-pm} with the `dispersion law' 
\begin{equation}
  \Theta_{h}\left(z,\bar{z}\right) =  -hz +  h^{-1}\bar{z} 
\end{equation}  
while the phases $\phi_{\pm}$ are given by 
\begin{equation}
 \phi_{+}\left(z,\bar{z}\right) = - \bar{z}/\zeta, 
 \qquad
 \phi_{-}\left(z,\bar{z}\right) = \zeta z.
\label{def-phi-pm}
\end{equation}  
The above formulae for the Toeplitz solutions demonstrate the typical for 
integrable systems phenomenon: solutions for the \textit{nonlinear} equations 
are determinants of matrices that satisfy \textit{linear} ones. This rule in 
our case can be extended as follows: the relationships between the 
$\tau$-functions $\mytau0_{n}$ and $\mytau1_{n}$ 
(which are, recall, solutions for the 2DTL used to construct $u_{n}$) 
become \textit{linear} when rewritten in terms of the `inside-determinant' 
objects, $\alpha_{k}$ and $\alpha_{k}^{\pm}$. 

As in the case of the infinite chain, let us consider one of the simplest 
solutions for \eref{alpha-semi}. Using the polar coordinates $z=re^{i\theta}$ 
one can obtain 
\begin{equation}
  \alpha_{k} = \exp\left[ - i (\theta + \pi/2) k \right] I_{k}(2r) 
\end{equation}  
where $I_{k}$ are the modified Bessel functions. 
This leads, together with definition \eref{def-phi-pm} of $\phi_{\pm}$, 
\begin{equation}
  \phi_{\pm}(r,\theta) = 
  \mp r \zeta^{\mp 1} e^{ \mp i \theta }, 
\end{equation}  
to 
\begin{equation}
  \alpha_{k}^{\pm} = 
  \exp\left[ - i (\theta + \pi/2) k \right]  
  \left[ I_{k}(2r) + \frac{ \phi_{\pm}(r,\theta) }{ ir } I_{k\pm 1}(2r) \right] 
\end{equation}  
and consequently to
\begin{equation}
  u_{n} = 
  u_{*} e^{i \phi_{\pm}(r,\theta) } 
  \frac{ \det\left| 
          I_{k+a-b}(2r) + \frac{ \phi_{\pm}(r,\theta) }{ ir } I_{k+a-b \pm 1}(2r) 
          \right|_{a,b=1,...,n} } 
       { \det\left| I_{k+a-b}(2r) \right|_{a,b=1,...,n} } 
\end{equation}
In context of the Heisenberg-like model \cite{PV11} these solutions describe 
circular magnetic domain structures.

\subsection{Finite chain.} 

The semi-infinite solutions derived in the previous subsection can be easily 
modified to provide the finite ones with $u_{n} \ne 0$ for $n=0,1,...,N$ 
only. To this end one has to define 
\begin{equation}
  u_{0}(z, \bar{z}) = u_{*} e^{ i\phi_{\pm}(z, \bar{z}) }, 
\label{def-uz}
\end{equation}
where $\phi_{\pm}$ are defined by \eref{def-phi-pm}, and to replace the integrals 
in \eref{sol-alpha} and \eref{sol-alpha-pm} with finite sums, 
\numparts
\begin{eqnarray}
  \alpha_{k}\left(z,\bar{z}\right) 
  & = & 
  \sum_{p=1}^{N} \hat\alpha_{p} \; 
  h_{p}^{k} 
  \exp\left[ - i \Theta_{h_{p}}\left(z,\bar{z}\right) \right]
\label{sol-alpha-fin}
\\
  \alpha_{k}^{\pm}\left(z,\bar{z}\right) 
  & = & 
  \sum_{p=1}^{N} \, \hat\alpha_{p} 
  \left[ 1 - \left(h_{p}/\zeta\right)^{\pm 1} \right] 
  h_{p}^{k} 
  \exp\left[ - i \Theta_{h_{p}}\left(z,\bar{z}\right) \right]
\label{sol-alpha-pm-fin}
\end{eqnarray}
\endnumparts
where
\begin{equation}
  \Theta_{h_{p}}\left(z,\bar{z}\right) =  -h_{p}z +  h_{p}^{-1}\bar{z}. 
\label{def-Theta-fin} 
\end{equation}  
These modifications do not change the fact that 
$i \partial \alpha_{k} = -\alpha_{k+1}$ and 
$i \bar\partial \alpha_{k} = \alpha_{k-1}$ 
which implies that $u_{0}$ and $u_{n}$ given by \eref{u-sol-semi} still solve 
\eref{se} for $n=1,...,N-1$. Thus, it remains to prove that they solve
\eref{se} for $n=0,N$ as well. The case $n=0$ is trivial because for both 
choices of $\phi_{\pm}$ 
\begin{equation}
  \partial\bar\partial u_{0} = 
  (\partial u_{0})(\bar\partial u_{0}) = 0 
\end{equation}
converting equation \eref{se} with $n=0$ into trivial one. 
Considering the right-end equation, one can show, using the identity 
\begin{equation}
  \det\left| \sum_{p=1}^{N} \alpha_{p} x_{p}^{a-b} \right|_{a,b=1,...,N} 
  = 
  (-1)^{\frac{1}{2} N(N-1)} 
  \prod_{p=1}^{N} \frac{ \alpha_{p} }{ x_{p}^{N-1} } 
  \prod_{1 \le p < q \le N} \left( x_{p} - x_{q} \right)^{2}, 
\end{equation}
that 
\begin{equation}
  u_{N}(z, \bar{z}) = u_{*}' e^{ i\phi_{\pm}(z, \bar{z}) }, 
\end{equation}
where $u_{*}'$ is a constant: the factors 
$\exp\left[ - i \Theta_{h_{p}}\left(z,\bar{z}\right) \right]$ cancel themselves 
when we divide $\det\left| \alpha^{\pm}_{k+a-b} \right|$ by 
$\det\left| \alpha_{k+a-b} \right|$ leading to 
\begin{equation}
  u_{*}' = u_{*} 
  \prod_{p=1}^{N} \left[ 1 - ( h_{p} / \zeta )^{\pm 1} \right] 
\end{equation}
Hence, 
\begin{equation}
  \partial\bar\partial u_{N} = 
  (\partial u_{N})(\bar\partial u_{N}) = 0, 
\end{equation}
ensuring solution of \eref{se} for $n=N$.

To summarize, $N+1$ functions, $u_{0}$ given by \eref{def-uz} and 
\begin{equation}
  u_{n}\left(z,\bar{z}\right) = 
  u_{0}\left(z,\bar{z}\right) \;
  \frac{ 
    \det\left| \alpha_{k+a-b}^{\pm}\left(z,\bar{z}\right) \right|_{a,b=1,...,n} 
  }{  
    \det\left| \alpha_{k+a-b}\left(z,\bar{z}\right) \right|_{a,b=1,...,n} 
  }, 
  \qquad n=1,...,N 
\end{equation}
with \eref{sol-alpha-fin}--\eref{def-Theta-fin} solve the system of $N+1$ 
equations \eref{se} for $n=0,...,N$.

\section{Soliton solutions. \label{sec-dark}}

The dark-soliton solutions for our equations can be constructed of the 
determinants 
\begin{equation}
  \myomega\left( \mymatrix{A} \right) 
  = 
  \det \left| \mymatrix{1} + \mymatrix{A} \right|
\end{equation}
of the $N \times N$ matrices $\mymatrix{A}$ that satisfy the `almost rank-1' 
condition 
\begin{equation}
  \mymatrix{L} \mymatrix{A} 
  - 
  \mymatrix{A} \mymatrix{R}  
  = 
  | \,\ell\, \rangle \langle a |. 
\label{rank-one}
\end{equation}
Here $\mymatrix{1}$ is the $N \times N$ unit matrix, 
$\mymatrix{L}$ and $\mymatrix{R}$ are the constant diagonal matrices, 
$| \,\ell\, \rangle$ is the constant $N$-component column, 
$| \,\ell\, \rangle = \left( \ell_{1}, ... , \ell_{N} \right)^{T}$
and $\langle a |$ is the $N$-component row depending on the coordinates,
$\langle a\left( z, \bar{z} \right) | = 
\left( 
  a_{1}\left( z, \bar{z} \right), ... , 
  a_{N}\left( z, \bar{z} \right) 
\right)$. 
The second part of the `solitonic ansatz', except for \eref{rank-one}, is that 
the dependence of $\mymatrix{A}$ on all coordinates ($z$, $\bar{z}$ and $n$) 
can be built by means of the shifts
\begin{equation}
  \mathbb{T}_{\zeta} \myomega = 
  \myomega\left( \mymatrix{A} \mymatrix{H}_{\zeta} \right)
\end{equation}
with 
the matrices $\mymatrix{H}_{\zeta}$ being defined by 
\begin{equation}
  \mymatrix{H}_{\zeta} 
  = 
  \left( \mymatrix{L} - \zeta \right) 
  \left( \mymatrix{R} - \zeta \right)^{-1}
\end{equation}
(we do not write the unit matrix explicitly, so 
$\left( \mymatrix{L} - \zeta \right)$ 
stands for $\left( \mymatrix{L} - \zeta\mymatrix{1} \right)$ \textit{etc}).

The remarkable property of the above matrices is that the determinants
\begin{equation}
  \myomega_{\zeta} = \mathbb{T}_{\zeta}\myomega, 
\qquad
  \myomega_{\xi\eta} = \mathbb{T}_{\xi}\mathbb{T}_{\eta}\myomega 
\end{equation}
satisfy the Fay-like identity
\begin{equation} 
  (\xi   - \eta)  \, \myomega_{\zeta} \, \myomega_{\xi\eta} 
  + 
  (\eta  - \zeta) \, \myomega_{\xi}   \, \myomega_{\eta\zeta} 
  +  
  (\zeta - \xi)   \, \myomega_{\eta}  \, \myomega_{\zeta\xi} 
  = 0 
\label{fay-dark}
\end{equation}
(see, \textit{e.g.}, appendix of \cite{V12}).
Introducing the differential operators $\partial_{\xi}$ defined by 
\begin{equation}
  \mathbb{T}_{\xi}^{-1} \, \mathbb{T}_{\xi+\delta} \, \myomega = 
  \myomega + i\delta \, \partial_{\xi} \myomega + 
  O\left( \delta^{2} \right) 
\label{dark-def-x}
\end{equation}
one can derive from \eref{fay-dark} the differential Fay identities
\begin{equation}
  (\xi - \alpha) \; i D_{\xi} \, \myomega_{\alpha} \cdot \myomega
  = 
  \left( \mathbb{T}_{\xi}^{-1} \myomega_{\alpha} \right) 
  \left( \mathbb{T}_{\xi} \myomega \right) 
  -
  \myomega_{\alpha} \myomega
\end{equation}
and 
\begin{equation}
  (\xi - \alpha)(\xi - \beta) \; 
  i D_{\xi} \, \myomega_{\alpha} \cdot \myomega_{\beta}
  = 
  (\alpha - \beta) 
  \left[
    \left( \mathbb{T}_{\xi}^{-1} \myomega_{\alpha\beta} \right) 
    \left( \mathbb{T}_{\xi} \myomega \right) 
    -
    \myomega_{\alpha} \myomega_{\beta}
  \right].
\label{fay-dark-diff}
\end{equation}
Applying these identities two times, one arrives at
\begin{equation}
  \frac{1}{2} \; (\xi - \eta)^{2} \; 
  D_{\xi}D_{\eta} \, \myomega \cdot \myomega
  = 
  \myomega^{2} 
  - 
  \left( \mathbb{T}_{\xi} \mathbb{T}_{\eta}^{-1} \myomega \right) 
  \left( \mathbb{T}_{\xi}^{-1} \mathbb{T}_{\eta} \myomega \right) 
\end{equation}
from which it is easy to obtain solutions for the 2DTL by associating 
the $\partial$, $\bar\partial$ operators with 
$\partial_{\xi}$ and $\partial_{\eta}$ (with fixed $\xi$ and $\eta$) and 
introducing the $n$-variable by means of the powers of $\mathbb{T}_{\nu}$
\begin{equation}
  \myomega_{n} 
  = 
  \mathbb{T}_{\nu}^{n} \myomega 
\end{equation}
with 
\begin{equation}
  \mymatrix{H}_{\xi} 
  = 
  \mymatrix{H}_{\eta}\mymatrix{H}_{\nu}. 
\label{dark-restr}
\end{equation}
From the Fay identities \eref{fay-dark} and \eref{fay-dark-diff} one can 
derive after straightforward calculations that the functions 
\numparts
\begin{eqnarray}
  \mypretau0_{n} & = & 
  \mypretau0_{*} \, h_{\alpha}^{n} \; \mathbb{T}_{\nu}^{n} \, \myomega_{\alpha} 
  \\[2mm] 
  \mypretau1_{n} & = & 
  \mypretau1_{*} \, h_{\beta}^{n} \; \mathbb{T}_{\nu}^{n} \, \myomega_{\beta} 
  \\[2mm] 
  \mypretau4_{n} & = & 
  \mypretau4_{*} \, \check{h}^{n} \; \mathbb{T}_{\nu}^{n} \, \mathbb{T}_{\eta}^{-1}
  \myomega_{\alpha\beta} 
  \\[2mm] 
  \mypretau3_{n} & = & 
  \mypretau3_{*} \, \hat{h}^{n} \; \mathbb{T}_{\nu}^{n} \, \myomega_{\xi} 
\end{eqnarray} 
\endnumparts
where the $h$-factors are defined by 
\begin{equation}
  h_{\alpha} = \frac{ \xi - \alpha }{ \eta - \alpha }, 
  \qquad
  h_{\beta} = \frac{ \xi - \beta }{ \eta - \beta },  
\end{equation}
\begin{equation}
  \check{h} = \mu \, h_{\alpha} h_{\beta} \, (\xi-\eta), 
  \qquad 
  \hat{h} = \bar\mu \, (\xi-\eta) 
\end{equation}
(with constants $\mu$ and $ \bar\mu$ satisfying $\mu\bar\mu = (\xi-\eta)^{-2}$) 
and the constants 
$\mypretau0_{*}$, $\mypretau1_{*}$, $\mypretau4_{*}$, $\mypretau3_{*}$ are related by 
\begin{equation}
  \mypretau4_{*} \mypretau3_{*} = 
  \mypretau0_{*} \mypretau1_{*} \; 
  \frac{ (\beta - \alpha)(\xi - \eta) }{ (\alpha - \eta)(\beta - \eta) } 
\end{equation}
satisfy the following equations:
\numparts
\begin{eqnarray}
\label{pre-dark-a}
  \left[ i D_{\xi} + \lambda_{\xi}(\beta,\alpha) \right]
  \mypretau1_{n} \cdot \mypretau0_{n} 
  & = & 
  \phantom{-}\mu \, \mypretau4_{n-1} \, \mypretau3_{n} 
  \\[2mm] 
  \left[ i D_{\xi} + \lambda_{\xi}(\beta,\eta) \right]
  \mypretau4_{n} \cdot \mypretau0_{n} 
  & = & 
  - \mu \, \mypretau4_{n-1} \, \mypretau0_{n+1} 
  \\[2mm] 
  \left[ i D_{\xi} + \lambda_{\xi}(\beta,\eta) \right]
  \mypretau1_{n} \cdot \mypretau3_{n-1} 
  & = & 
  - \mu \, \mypretau1_{n-1} \, \mypretau3_{n} 
\end{eqnarray}
\endnumparts
and 
\numparts
\begin{eqnarray}
  \left[ i D_{\eta} + \lambda_{\eta}(\beta,\alpha) \right]
  \mypretau1_{n} \cdot \mypretau0_{n} 
  & = & 
  \bar\mu \, \mypretau3_{n-1} \, \mypretau4_{n} 
  \\[2mm] 
  \left[ i D_{\eta} + \lambda_{\eta}(\beta,\xi) \right]
  \mypretau4_{n-1} \cdot \mypretau0_{n} 
  & = & 
  \bar\mu \, \mypretau0_{n-1} \, \mypretau4_{n} 
  \\[2mm] 
  \left[ i D_{\eta} + \lambda_{\eta}(\beta,\xi) \right]
  \mypretau1_{n} \cdot \mypretau3_{n} 
  & = & 
  \bar\mu \, \mypretau3_{n-1} \, \mypretau1_{n+1}. 
\label{pre-dark-f}
\end{eqnarray}
\endnumparts
Here 
\begin{equation}
  \lambda_{\xi}(\alpha,\beta) 
  = 
  \lambda_{\xi}(\alpha) - \lambda_{\xi}(\beta) 
\end{equation}
with 
\begin{equation}
  \lambda_{\xi}(\gamma) = \frac{ 1 }{ \xi - \gamma }. 
\end{equation}

Comparing the above equations with \eref{main-pos} and 
\eref{main-neg} it is easy to conclude that 
to obtain solutions for our problem one has to identify 
\begin{equation}
  \partial = \mu^{-1} \partial_{\xi} 
  \hspace{10mm}
  \bar\partial = - \bar\mu^{-1} \partial_{\eta} 
\label{dark-dbd}
\end{equation}
and to introduce some linear in $z$ and $\bar{z}$ phases to eliminate extra 
$\lambda$-terms in \eref{pre-dark-a}--\eref{pre-dark-f}. This leads to the 
following expression for $u_{n}$:
\begin{equation}
  u_{n}\left(z,\bar{z}\right) = 
  u_{*} 
  e^{ i\phi\left(z,\bar{z}\right) } \, 
  h^{n} \, 
  \frac{\det\left| 
           \mymatrix{1} 
           + \mymatrix{A}_{n}\left(z,\bar{z}\right)\mymatrix{H}_{\beta} 
  \right|}
       {\det\left| 
          \mymatrix{1} 
          + \mymatrix{A}_{n}\left(z,\bar{z}\right)\mymatrix{H}_{\alpha} 
  \right|}
\label{dark-u}
\end{equation}
where 
\begin{equation}
  h = \frac{ h_{\beta} }{ h_{\alpha} }, 
\end{equation}
\begin{equation}
  \mymatrix{A}_{n}\left(z,\bar{z}\right) 
  = 
  \mymatrix{A}\left(z,\bar{z}\right) \, \mymatrix{H}_{\nu}^{n} 
\end{equation}
and 
\begin{equation}
  \phi = 
  \mu^{-1} \, \lambda_{\xi}(\alpha,\beta) z 
  + 
  \bar\mu^{-1} \, \lambda_{\eta}(\beta,\alpha) \bar{z} 
  + \mbox{constant}. 
\end{equation}
Finally, it remains to resolve the restriction \eref{dark-restr} and to write 
down explicitly the dependence of $\mymatrix{A}$ on the coordinates, which 
can be done by calculating the limit in \eref{dark-def-x}, 
\begin{equation}
  i \mymatrix{A}^{-1}\partial_{\zeta} \mymatrix{A} = 
  \left( \mymatrix{L} - \mymatrix{R} \right) 
  \left( \mymatrix{L} - \zeta \right)^{-1}
  \left( \mymatrix{R} - \zeta \right)^{-1} 
\end{equation}
(for any $\zeta$).

The restriction \eref{dark-restr} implies that the matrices $\mymatrix{L}$ and 
$\mymatrix{R}$ are not independent, 
\begin{equation}
  \left( \mymatrix{L} - \xi \right) 
  \left( \mymatrix{R} - \xi \right) 
  =  
  ( \nu - \xi )( \eta - \xi ) \; \mymatrix{1}. 
\end{equation}
Introducing matrices $\hat\mymatrix{L}$ and $\hat\mymatrix{R}$,  
\begin{equation}
  \hat\mymatrix{L} 
  = 
  \frac{ 1 }{ \xi - \nu } 
  \left( \mymatrix{L} - \nu \right), 
  \qquad 
  \hat{R} 
  = 
  \frac{ 1 }{ \xi - \nu } 
  \left( \mymatrix{R} - \nu \right) 
\end{equation}
satisfying 
\begin{equation}
  \left( \hat\mymatrix{L} - \mymatrix{1} \right) 
  \left( \hat\mymatrix{R} - \mymatrix{1} \right) 
  = 
  \frac{ \xi - \eta }{ \xi - \nu } \, 
  \mymatrix{1} 
\end{equation}
and parameter $f$, instead of $\mu$ and $\bar\mu$,
\begin{equation}
  f = \frac{ 1 }{ \mu (\xi - \eta) } 
  = \bar\mu ( \xi - \eta )  
\end{equation}
the $z$- and $\bar{z}$-dependence of $\mymatrix{A}$ and $\phi$ can be presented 
in a symmetric form as 
\begin{equation}
  \mymatrix{A}\left(z,\bar{z}\right) 
  = 
  \mymatrix{A}_{*} 
  \exp\left\{  i\mymatrix{M}\left(z,\bar{z}\right) \right\}  
\end{equation}
where $\mymatrix{A}_{*}$ is a constant matrix,   
\begin{equation}
  \mymatrix{M}\left(z,\bar{z}\right) = 
  f \left( \hat\mymatrix{R} - \hat\mymatrix{L} \right) z
  + 
  f^{-1} \left( \hat\mymatrix{R}^{-1} - \hat\mymatrix{L}^{-1} \right) \bar{z} , 
\end{equation}
and 
\begin{equation}
  \phi\left(z,\bar{z}\right)
  =  
  f \left( h_{\beta}^{-1} - h_{\alpha}^{-1} \right) z 
  + 
  f^{-1} \left( h_{\beta} - h_{\alpha} \right) \bar{z}. 
\end{equation}
These formulae together with \eref{dark-u} describe the $N$-dark-soliton 
solutions for our problem.

The simplest (one-soliton) solution can be written as
\begin{equation}
  u_{n}(z, \bar{z}) = 
  u_{*}
  e^{ 2i \varphi_{n}(z, \bar{z}) } 
  \left[ 1 + \tanh\rho \; \tanh U_{n}(z, \bar{z}) \right] 
\end{equation}
where 
\begin{equation}
	\rho = 
	\frac{ 1 }{ 2 } \ln
	\frac{ 1 - h_{\alpha}\hat{L} }{ 1 - h_{\alpha}\hat{R} } \,
	\frac{ 1 - h_{\beta}\hat{R} }{ 1 - h_{\beta}\hat{L} },
\end{equation}
(note that now $\hat{L}$ and $\hat{R}$ are just complex numbers)  
the phase $\varphi_{n}$ is given by 
\begin{equation}
  \varphi_{n}(z, \bar{z}) = \xi_{0}z + \eta_{0}\bar{z} + \zeta_{0} n 
\end{equation}
with
\begin{equation}
	\xi_{0} = 
	\frac{ f }{ 2 } 
	\left( h_{\beta}^{-1} - h_{\alpha}^{-1}  \right), 
  \qquad
	\eta_{0} = 
	\frac{ 1 }{ 2f } 
	\left( h_{\beta} - h_{\alpha}  \right) 
\end{equation}
and 
\begin{equation}
  \sin^{2} \zeta_{0} = \xi_{0} \eta_{0}, 
\end{equation}
while $U_{n}$ is given by 
\begin{equation}
  U_{n}(z, \bar{z}) = \xi_{s}z + \eta_{s}\bar{z} + \zeta_{s} n 
\end{equation}
with
\begin{equation}
	\xi_{s} = 
	\frac{ if }{ 2 } 
	\left( \hat{R} - \hat{L} \right), 
  \qquad
	\eta_{s} = 
	\frac{ i }{ 2f } 
  \left( \hat{R}^{-1} - \hat{L}^{-1} \right) 
\end{equation}
and 
\begin{equation}
  \sinh^{2} \zeta_{s} = \xi_{s} \eta_{s}. 
\end{equation}

To conclude, we would like to note that in the soliton case 
the link between $\mytau0_{n}$ and $\mytau1_{n}$ is \textit{linear} in the 
terms of the matrices $\mymatrix{A}_{n}$: the matrices that appear in the 
determinants in \eref{dark-u} are related by the constant diagonal matrix 
$\mymatrix{H}_{\beta}\mymatrix{H}_{\alpha}^{-1}$.

\section{Quasiperiodic solutions. \label{sec-qps}}

In this section, we derive the periodic solutions for our equation proceeding in 
the way similar to the one used in the previous section. 
The main difference is in the starting point: instead of identity 
\eref{fay-dark} we use the original Fay's trisecant identity 
(see \eref{fay} below).

The solutions that we derive below are combinations of the $\theta$-functions 
defined over a compact Riemann surface $\Gamma$ of the genus $g$ for which 
one can choose in a standard way a set of 
closed contours (cycles) $\{ a_{i}, b_{i} \}_{i=1,..., g}$ with the intersection 
indices
\begin{equation}
a_{i} \circ a_{j} =
b_{i} \circ b_{j} = 0,
\qquad
a_{i} \circ b_{j} = \delta_{ij}
\qquad
i,j = 1, \dots, g
\end{equation}
and $g$ independent holomorphic differentials 
$\varpi_{k}$ 
satisfying the normalization conditions
\begin{equation}
  \oint_{a_{i}} \varpi_{k} = \delta_{ik},
  \qquad
  i,k = 1, \dots, g.
\end{equation}
The matrix of the $b$-periods,
\begin{equation}
  \Omega_{ik} = \oint_{b_{i}} \varpi_{k},
  \qquad
  i,k = 1, \dots, g 
\end{equation}
determines the so-called period lattice,
$L_{\Omega} = \left\{
   \vec{m} + \Omega \vec{n},
   \quad
   \vec{m}, \vec{n} \in {\mathbb Z}^{g}
   \right\}$,
and the Abel mapping from $\Gamma$ to the 2$g$ torus 
${\mathbb C}^{g}/L_{\Omega}$ (the Jacobian of this surface),
\begin{equation}
  P \to \int^{P}_{P_{0}} \vec{\omega}
\label{abel}
\end{equation}
where $P$ is a point of $\Gamma$,
$\vec{\omega}$ is the $g$-vector of the 1-forms,
$\vec{\omega} =
\left( \varpi_{1}, \dots, \varpi_{g} \right)^{\scriptscriptstyle T}$, 
and $P_{0}$ is some fixed point of $\Gamma$.

The  $\theta$-function, $\theta(\vec{\zeta})=\theta(\vec{\zeta},\Omega)$, 
is defined by 
\begin{equation}
\theta\left(\vec{\zeta}\right) =
\sum_{ \vec{n} \, \in \, \mathop{\mathbb Z}\nolimits^{g} }
\exp\left\{
     \pi i \, \biggl( \vec{n}, \Omega \vec{n} \biggr) \; +
   2 \pi i \, \biggl( \vec{n}, \vec{\zeta} \biggr)
\right\}
\end{equation}
where $( \vec{n},\vec{\zeta} )$ stands for $\sum_{i=1}^{g} n_{i}\zeta_{i}$. 
This is a quasiperiodic function on $\mathbb{C}^{g}$
\numparts
\begin{eqnarray}
\theta\left(\vec{\zeta} + \vec{n}\right) &=&
   \theta\left(\vec{\zeta}\right)
\\
\theta\left(\vec{\zeta} + \Omega\vec{n}\right) &=&
   \exp\left\{
      -   \pi i \, \biggl( \vec{n}, \Omega \vec{n} \biggr) \;
      - 2 \pi i \, \biggl( \vec{n}, \vec{\zeta} \biggr)
   \right\}
\theta\left(\vec{\zeta}\right)
\end{eqnarray}
\endnumparts
for any $\vec{n} \in {\mathbb Z}^{g}$.

The calculations presented below are based on the famous Fay's trisecant formula 
\cite{Fay,Mumford2} that can be written as
\begin{equation}
  \varepsilon^{P}_{B} \, 
  \varepsilon^{Q}_{A} \; 
  \theta^{P}_{A} \, 
  \theta^{Q}_{B} 
- 
  \varepsilon^{P}_{A} \, 
  \varepsilon^{Q}_{B} \; 
  \theta^{P}_{B} \, 
  \theta^{Q}_{A} 
+ 
  \varepsilon^{P}_{Q} \, 
  \varepsilon^{A}_{B} \; 
  \theta \, 
  \theta^{P Q}_{A B} 
= 0.
\label{fay}
\end{equation}
Here
\begin{equation}
  \theta^{Q_{1}...Q_{m}}_{P_{1}...P_{m}} = 
  \theta\left( 
    \vec{\zeta} 
    + \sum_{i=1}^{m}\int\nolimits^{Q_{i}}_{P_{i}} \vec{\omega} 
    \right) 
\end{equation}
and the skew-symmetric function $\varepsilon^{Q}_{P}$, 
$\varepsilon^{Q}_{P}=-\varepsilon^{P}_{Q}$, 
is given by
\begin{equation}
  \varepsilon^{Q}_{P} = 
  \theta\left( \vec{e} + \int\nolimits^{Q}_{P} \vec{\omega} \right)
\end{equation}
where $\vec{e}$ is a zero of the $\theta$-function: $\theta\left(\vec{e}\right)=0$.

Now let us define the differential operators $\partial_{X}$ by 
\begin{equation}
  \theta^{P}_{X} = 
  \theta 
  + \varepsilon^{P}_{X} \, \partial_{X}\theta 
  + o\left( \varepsilon^{P}_{X} \right). 
\end{equation}
In what follows, we use $\partial_{A}$ and $\partial_{B}$ defined near two 
points $A$ and $B$ that are fixed. One of them will play the role of 
$\partial$ and another of $\bar\partial$. 
Taking the limit in \eref{fay} one can obtain the differential Fay's identity 
\begin{equation}
  \left[ D_{X} + \lambda_{X}(P,Q) \right] 
  \theta^{P}_{Q} \cdot \theta 
  = 
  \gamma_{X}(P,Q) \; 
  \theta^{P}_{X} \, \theta^{X}_{Q} 
\label{fay-diff}
\end{equation}
where $D_{X}$ is the Hirota operator corresponding to $\partial_{X}$ and 
the functions $\lambda_{X}(P)$ and $\gamma_{X}(P,Q)$ are defined by 
\begin{equation}
  \lambda_{X}(P,Q) = \lambda_{X}(P) - \lambda_{X}(Q) 
\end{equation}
with 
\begin{equation}
  \lambda_{X}(P) = 
  \lim_{Y \to X} 
  \frac{ \varepsilon^{P}_{Y} - \varepsilon^{P}_{X} }
       { \varepsilon^{Y}_{X} \, \varepsilon^{P}_{X} }
\end{equation}
and 
\begin{equation}
  \gamma_{X}(P,Q) = 
  \frac{ \varepsilon^{P}_{Q} }{ \varepsilon^{P}_{X} \, \varepsilon^{Q}_{X} }. 
\end{equation}

Consider the functions 
\numparts
\begin{eqnarray}
\label{qps-pre-a}
	\mypretau0_{n} & = & 
	\mypretau0_{*} \left[h(P)\right]^{n} 
  \theta\!\left( \vec{\zeta}_{n} + \int\nolimits^{P}_{A} \vec{\omega} \right) 
  \\[1mm]
	\mypretau1_{n} & = & 
	\mypretau1_{*} \left[h(Q)\right]^{n} 
  \theta\!\left( \vec{\zeta}_{n} + \int\nolimits^{Q}_{A} \vec{\omega} \right) 
  \\[1mm]
\label{qps-pre-c}
	\mypretau4_{n} & = & 
	\mypretau4_{*} \check{h}^{n} \; 
  \theta\!\left( \vec{\zeta}_{n} + \int\nolimits^{PQ}_{AB} \vec{\omega} \right) 
  \\[1mm]
	\mypretau3_{n} & = & 
	\mypretau3_{*} \hat{h}^{n} \;
  \theta\!\left( \vec{\zeta}_{n} \right) 
\end{eqnarray}
\endnumparts
where 
\begin{equation}
  h(P) = \frac{ \varepsilon^{P}_{A} }{ \varepsilon^{P}_{B} } 
\end{equation}
and the $n$-dependence is given by 
\begin{equation}
  \vec{\zeta}_{n+1} 
  = 
  \vec{\zeta}_{n} + \int\nolimits^{A}_{B} \vec{\omega}. 
\end{equation}
It follows from \eref{fay} that if one imposes the restrictions 
\begin{equation}
	\mypretau4_{*} \mypretau3_{*} =  
	\mypretau0_{*} \mypretau1_{*} \; 
  \frac{ \varepsilon^{P}_{Q} \, \varepsilon^{B}_{A} }
       { \varepsilon^{P}_{B} \, \varepsilon^{Q}_{B} }
\end{equation}
and 
\begin{equation}
	\check{h}\hat{h} = h(P)h(Q), 
\end{equation}
then these functions satisfy 
\begin{equation}
	\mypretau0_{n} \mypretau1_{n+1} 
	- 
	\mypretau0_{n+1} \mypretau1_{n} 
	= 
	\mypretau4_{n} \mypretau3_{n}. 
\end{equation}
Furthermore, by taking 
\begin{equation}
	\check{h} = \mu_{A} \varepsilon^{B}_{A}  
	\; 
  \frac{ \varepsilon^{P}_{A} \, \varepsilon^{Q}_{A} }
       { \varepsilon^{P}_{B} \, \varepsilon^{Q}_{B} },
\qquad
	\hat{h} = \frac{ 1 }{ \mu_{A} \varepsilon^{B}_{A} } 
\end{equation}
where $\mu_{A}$ and $\mu_{B}$ are two constants related by 
\begin{equation}
	\mu_{A} \mu_{B} \left( \varepsilon^{B}_{A} \right)^{2} = 1 
\end{equation}
one can get from \eref{fay-diff} the set of identities which will be 
associated with the $\partial$-flows:
\numparts
\begin{eqnarray}
\label{per-pre-a}
  \left[ D_{A} + \lambda_{A}(Q,P) \right] 
  \mypretau1_{n} \cdot \mypretau0_{n} 
  & = & 
  - \mu_{A} \, \mypretau4_{n-1} \, \mypretau3_{n}
  \\[2mm] 
  \left[ D_{A} + \lambda_{A}(Q,B) \right] 
  \mypretau4_{n} \cdot \mypretau0_{n} 
  & = & 
  \phantom{-}\mu_{A} \, \mypretau4_{n-1} \, \mypretau0_{n+1}
  \\[2mm] 
  \left[ D_{A} + \lambda_{A}(Q,B) \right] 
  \mypretau1_{n} \cdot \mypretau3_{n-1} 
  & = & 
  \phantom{-}\mu_{A} \, \mypretau1_{n-1} \, \mypretau3_{n} 
\end{eqnarray}
\endnumparts
and another one, 
\numparts
\begin{eqnarray}
  \left[ D_{B} + \lambda_{B}(Q,P) \right] 
  \mypretau1_{n} \cdot \mypretau0_{n} 
  & = & 
  - \mu_{B} \, \mypretau3_{n-1} \, \mypretau4_{n} 
  \\[2mm] 
  \left[ D_{B} + \lambda_{B}(Q,A) \right] 
  \mypretau4_{n-1} \cdot \mypretau0_{n} 
  & = & 
  - \mu_{B} \, \mypretau0_{n-1} \, \mypretau4_{n} 
  \\[2mm] 
  \left[ D_{B} + \lambda_{B}(Q,A) \right] 
  \mypretau1_{n} \cdot \mypretau3_{n} 
  & = & 
  - \mu_{B} \, \mypretau3_{n-1} \, \mypretau1_{n+1}
\label{per-pre-f}
\end{eqnarray}
\endnumparts
associated with the $\bar\partial$-equations. 

Comparing the above equations with \eref{main-pos} and \eref{main-neg}, one 
arrives at 
\begin{equation}
  \partial = \frac{ i }{ \mu_{A} } \, \partial_{A}, 
  \qquad
  \bar\partial = \frac{ 1 }{ i \mu_{B} } \, \partial_{B} 
\end{equation}
which determines the dependence on $z$ and $\bar{z}$, 
\begin{equation}
  \vec{\zeta}_{n}\left(z,\bar{z}\right) = 
  \vec{\zeta}_{*} 
  + z \vec{a} 
  + \bar{z} \vec{b} 
  + n \vec{c}
\end{equation}
where $\vec{c}$ was defined above, 
\begin{equation}
  \vec{c} = \int\nolimits^{A}_{B} \vec{\omega}, 
\end{equation}
while
\numparts
\begin{eqnarray}
  \vec{a} & = & 
  i f 
  \lim_{P \to A} \; 
  \frac{ \varepsilon^{P}_{B} }{ \varepsilon^{P}_{A} } \,
  \int\nolimits^{P}_{A} \vec{\omega} 
  \\[2mm]
  \vec{b} & = & 
  i f^{-1} 
  \lim_{P \to B} \; 
  \frac{ \varepsilon^{P}_{A} }{ \varepsilon^{P}_{B} } \,
  \int\nolimits^{P}_{B} \vec{\omega} 
\end{eqnarray}
\endnumparts
where we have introduced the constant $f$ instead of $\mu_{A}$ and $\mu_{B}$, 
\begin{equation}
  f = 
  \frac{ 1 }{ \mu_{A} \, \varepsilon^{A}_{B} } 
  = 
  \mu_{B} \, \varepsilon^{A}_{B}. 
\end{equation}
Again, the $\lambda$-terms in \eref{per-pre-a}--\eref{per-pre-f} can be 
eliminated by adding linear in $z$ and $\bar{z}$ phases, which leads to 
\begin{equation}
  u_{n} = \frac{ \mypretau1_{n} }{ \mypretau0_{n} } \, e^{ i\phi } 
\end{equation}
where
\begin{equation}
  \phi\left(z,\bar{z}\right) =  
  f \, \varepsilon^{B}_{A}\lambda_{A}(P,Q) \, z
  + 
  f^{-1} \varepsilon^{A}_{B}\lambda_{B}(P,Q) \, \bar{z} 
  + 
  \mbox{constant}. 
\end{equation}
Using the definitions of $\mypretau1_{n}$ and $\mypretau0_{n}$, and 
introducing $h = h(Q)/h(P)$,  
\begin{equation}
  h  = 
  \frac{ \varepsilon^{Q}_{A}\varepsilon^{P}_{B} }
       { \varepsilon^{P}_{A}\varepsilon^{Q}_{B} }, 
\end{equation}
we can write the final expression for the quasiperiodic solutions as 
\begin{equation}
  u_{n}\left(z,\bar{z}\right) = 
  u_{*}
  e^{ i\phi\left(z,\bar{z}\right) } \; 
  h^{n} \; 
  \frac{ 
    \theta\!\left( 
      \vec{\zeta}_{n}\left(z,\bar{z}\right) 
      + \int\nolimits^{Q}_{A} \vec{\omega} 
    \right) 
  }{  
    \theta\!\left( 
      \vec{\zeta}_{n}\left(z,\bar{z}\right) 
      + \int\nolimits^{P}_{A} \vec{\omega} 
    \right) 
  }.
\label{qps-u}
\end{equation}
This time, the link between $\mytau1_{n}$ and $\mytau0_{n}$ is \textit{linear} 
in terms of $\vec{\zeta}_{n}$: the transformation 
$\mytau0_{n} \to \mytau1_{n}$ is achieved by 
$\vec{\zeta}_{n} \to \vec{\zeta}_{n} + \int\nolimits^{Q}_{P} \vec{\omega}$.

The simplest of the quasiperiodic solutions \eref{qps-u}, 
after some slight modifications, can be rewritten as a cnoidal wave: 
\begin{equation}
  u_{n}(z, \bar{z}) = 
  u_{*}
  e^{ i \phi_{n}(z, \bar{z}) } 
  \mathop{\mbox{sn}} \zeta_{n}(z, \bar{z}) 
\end{equation}
where the phase $\phi_{n}$ and the function $\zeta_{n}$ are given by 
\numparts
\begin{eqnarray}
  \phi_{n}(z, \bar{z}) & = & \xi_{0}z + \eta_{0}\bar{z} + \delta_{0} n, 
\\
  \zeta_{n}(z, \bar{z}) & = & \xi_{p}z + \eta_{p}\bar{z} + \delta_{p} n 
\end{eqnarray}
\endnumparts
and $\mathop{\mbox{sn}} z = \mathop{\mbox{sn}}( z, k )$ is the elliptic sine. 
The parameters $\xi_{0,p}$, $\eta_{0,p}$, $\delta_{0,p}$ and the elliptic 
modulus $k$ are related by 
\numparts
\begin{eqnarray}
&&
  \xi_{p}\eta_{p} 
  = 
  \mathop{\mbox{sn}}\nolimits^{2} \delta_{p} 
\\
&&
  \xi_{0}\eta_{p} + \xi_{p}\eta_{0} 
  = 
  2 \, \sin \delta_{0} \, \mathop{\mbox{sn}} \delta_{p} 
\end{eqnarray}
and 
\begin{equation}
  \xi_{0}\eta_{0} 
  = 
  \left| 
    \mathop{\mbox{dn}} \delta_{p} 
    - e^{ i\delta_{0} } \mathop{\mbox{cn}}\delta_{p}  
  \right|^{2} 
\end{equation}
\endnumparts
for real $\delta_{0}$ and $\delta_{p}$.

\section{Solutions of the two-dimensional Volterra equation. \label{sec-tdve}}

As was mentioned in section \ref{sec-etal}, the authors of \cite{LSS80,LSS81}
derived general solutions of the 2DVE in the case of finite chain. 
Here we would like to present several other classes of solutions for this system, 
namely ones that can be obtained from the results presented in this paper using 
\eref{ab-tau-a} and \eref{ab-tau-b}. 

\subsection{Toeplitz solutions.}

As follows from \eref{ab-tau-a} and \eref{ab-tau-b}, 
the constants $\mytau0_{*}$ and $\mytau4_{*}$ as well as the phases 
$\bar{z}/\zeta$ (that we defined in 
\eref{toep-inf-a}, \eref{toep-inf-c} and 
\eref{toep-semi-a}, \eref{toep-semi-c}) 
disappear from the final formulae for $a_{n}$ and $b_{n}$, 
that can be written as 
\begin{equation}
  a_{n} = 
  \frac{ \mydet{A}{n+1}{\ell} \mydet{B}{n-1}{\ell-1} }
       { \mydet{A}{n}{\ell} \mydet{B}{n}{\ell-1} },
\qquad
  b_{n} = 
  \frac{ \mydet{A}{n-1}{\ell} \mydet{B}{n}{\ell-1} }
       { \mydet{A}{n}{\ell} \mydet{B}{n-1}{\ell-1} }
\end{equation}
in the infinite case ($-\infty < n < \infty$) and  
\begin{equation}
  a_{n} = 
  \frac{ \mydet{A}{k}{n+1} \mydet{B}{k-1}{n-1} }
       { \mydet{A}{k}{n} \mydet{B}{k-1}{n} },
\qquad
  b_{n} = 
  \frac{ \mydet{A}{k}{n-1} \mydet{B}{k-1}{n} }
       { \mydet{A}{k}{n} \mydet{B}{k-1}{n-1} }
\end{equation}
in the semi-infinite case ($1 \le n < \infty$). 
One can easily obtain similar solutions with $\mydet{C}{n}{\ell}$- and 
$\mydet{C}{k}{n}$-determinants which we do not write here.

\subsection{Soliton solutions.}

Making the shift 
$\mymatrix{A}_{n} \to 
\mymatrix{A}_{n}\mymatrix{H}_{\xi}\mymatrix{H}_{\alpha}^{-1}$  
and introducing the matrix $\mymatrix{B}_{n}$, 
\begin{equation}
  \mymatrix{B}_{n} = \mymatrix{A}_{n}\mymatrix{H}_{\beta}
\end{equation}
one can presents soliton solutions for \eref{se-ab} in the following form: 
\numparts
\begin{eqnarray}
  a_{n} = 
  c_{*} \;
  \frac{
    \det\left| 
       \mymatrix{1} 
       + \mymatrix{A}_{n}\mymatrix{H}_{\xi}\mymatrix{H}_{\nu} 
    \right| \; 
    \det\left| 
       \mymatrix{1} 
       + \mymatrix{B}_{n} 
    \right|
  }{
    \det\left| 
       \mymatrix{1} 
       + \mymatrix{A}_{n}\mymatrix{H}_{\xi} 
    \right| \; 
    \det\left| 
       \mymatrix{1} 
       + \mymatrix{B}_{n}\mymatrix{H}_{\nu} 
    \right|
  }
\\[2mm]
  b_{n} = 
  \frac{1}{c_{*}} \;
  \frac{
    \det\left| 
       \mymatrix{1} 
       + \mymatrix{A}_{n}\mymatrix{H}_{\eta} 
    \right| \; 
    \det\left| 
       \mymatrix{1} 
       + \mymatrix{B}_{n}\mymatrix{H}_{\nu} 
    \right|
  }{
    \det\left| 
       \mymatrix{1} 
       + \mymatrix{A}_{n}\mymatrix{H}_{\eta}\mymatrix{H}_{\nu} 
    \right| \; 
    \det\left| 
       \mymatrix{1} 
       + \mymatrix{B}_{n} 
    \right|
  }
\end{eqnarray}
\endnumparts
with $c_{*} = f / h_{\beta}$ and 
matrices $\mymatrix{A}_{n}=\mymatrix{A}_{n}\left(z,\bar{z}\right)$ and 
$\mymatrix{H}_{\zeta}$ defined in section \ref{sec-dark}. 
In these expressions one can find a symmetry 
$1/b_{n} = a_{n}( \xi\!\to\!\eta )$ which is a manifestation of many symmetries 
inherent in the 2DVE.

\subsection{Periodic solutions.}

It follows from \eref{ab-tau-a} and \eref{ab-tau-b} that the phases that one 
has to introduce to eliminate the $\lambda$-terms in 
\eref{per-pre-a}--\eref{per-pre-f} cancel themselves and formulae 
\eref{qps-pre-a}, \eref{qps-pre-c} yield \textit{periodic} solutions for 
the 2DVE. 
After the shift 
$\vec{\zeta}_{n} \to \vec{\zeta}_{n} - \int\nolimits^{P}_{A} \vec{\omega}$ 
one can write these solutions as 
\numparts
\begin{eqnarray}
  a_{n} & = & 
  c_{*} \; 
  \frac{ 
    \theta\!\left( \vec{\zeta}_{n} + \int\nolimits^{A}_{B} \vec{\omega} \right) 
    \theta\!\left( \vec{\zeta}_{n} + \int\nolimits^{Q}_{A} \vec{\omega} \right) 
  }{
    \theta\!\left( \vec{\zeta}_{n} \right) 
    \theta\!\left( \vec{\zeta}_{n} + \int\nolimits^{Q}_{B} \vec{\omega} \right) 
  }
\\[2mm]
  b_{n} & = & 
  \frac{1}{c_{*}} \;
  \frac{ 
    \theta\!\left( \vec{\zeta}_{n} + \int\nolimits^{B}_{A} \vec{\omega} \right) 
    \theta\!\left( \vec{\zeta}_{n} + \int\nolimits^{Q}_{B} \vec{\omega} \right) 
  }{
    \theta\!\left( \vec{\zeta}_{n} \right) 
    \theta\!\left( \vec{\zeta}_{n} + \int\nolimits^{Q}_{A} \vec{\omega} \right) 
  }
\end{eqnarray}
\endnumparts
where $c_{*} = f \varepsilon^{B}_{Q} / \varepsilon^{Q}_{A}$
with $\vec{\zeta}_{n}\left(z,\bar{z}\right)$ and $f$ defined in section 
\ref{sec-qps}. 
One can see that of two points that parametrize solutions \eref{qps-u}, 
$P$ and $Q$, only one is left and, again, one can find the symmetry linking 
$a_{n}$ and $b_{n}$:
$b_{n} \propto a_{n}(A \!\leftrightarrow\! B)$.

\section{Conclusion}

We have obtained three types of explicit solutions for the model \eref{se-vect}. 
To conclude, we would like to discuss a few questions that have not been studied 
comprehensively in this paper. 
The first one is related to whether these solutions are real or complex.
All presented above are, in general, complex ones.
From the viewpoint of applications there are situations when namely such solutions 
are needed. 
For example, in the model of the graphite-like magnetics with Heisenberg-type 
interaction discussed in \cite{PV11} 
$u_{n} \propto \sigma_{n}^{(1)} + i \sigma_{n}^{(2)}$ 
where $\sigma_{n}^{(k)}$ are the components of 
$\vec{\sigma}_{n}$,  
$\vec{\sigma}_{n} = \left( \sigma_{n}^{(1)}, \sigma_{n}^{(2)}, \sigma_{n}^{(3)} \right)$, 
which means that in general situation $u_{n}$ is complex. 
However, sometimes one may be interested in the real solutions, as \textit{e.g.}, 
in the theory of hydrodynamic-type systems \cite{F97}. So, one faces a natural 
question of restrictions that should be imposed on the parameters of solutions 
that ensure their reality. This is a particular case of a problem which 
frequently arises in the theory of integrable equations, especially in the 
case of (quasi)periodic solutions, and surely deserves special studies.

Another range of questions is related to the symmetries of our problem. 
The symmetry $\delta \to - \delta$ in \eref{def-delta} and the following 
equations, which is another form of $n \to - n$ symmetry, is the simplest one. 
The symmetries that we saw in the previous section, 
$1/b_{n} = a_{n}( \xi\!\to\!\eta )$ and 
$b_{n} \propto a_{n}(A \!\leftrightarrow\! B)$ are less trivial. 
At the level of the ALH, they describe the links between various Miwa's shifts.
Since the ALH is out of the scope of this paper (though implicitly we used 
some of its properties), we do not discuss this topic further, noting only 
that studies in this direction may lead to some kind of nonlinear 
superposition formulae for our problem.

Finally, we would like to repeat that from the viewpoint of the 2DVE or RTC 
our main result, solutions of \eref{se-vect}, are nothing but solutions of the 
auxiliary linear problems \eref{rtc-sp}--\eref{rtc-evol-neg}. 
In other words, after some additional work (such as, \textit{e.g.}, extracting the 
the dependence on $\Lambda$) one can obtain from our results the Baker-Akhiezer 
function of the 2DVE or RTC, which is usually more complicated problem that to 
obtain solutions for an equation itself and which can be useful for applications.

\appendix 
\section{Toepllitz determinants.}

Applying the Jacobi identity 
\begin{equation}
  \Delta \, \Delta^{j_{1}j_{2}}_{k_{1}k_{2}} 
  = 
  \Delta^{j_{1}}_{k_{1}} \, \Delta^{j_{2}}_{k_{2}} 
  - 
  \Delta^{j_{1}}_{k_{2}} \, \Delta^{j_{2}}_{k_{1}} 
\label{det-jacobi}
\end{equation}
where $\Delta$ is the determinant of a matrix, $\Delta^{j}_{k}$ is the 
determinants of the same matrix with $j$th row and $k$th column being excluded, 
\textit{etc},  
to $\mydet{F}{k}{\ell+1}(\zeta)$ 
for $\left( j_{1},j_{2};k_{1},k_{2} \right)$ being equal to 
$\left( 1, \ell+1; 1, \ell + 1 \right)$, 
$\left( 1, 2; 1, \ell + 1 \right)$, 
$\left( 1, 3; 1, \ell + 1 \right)$ and  
$\left( 1,  \ell; 1, \ell + 1 \right)$ one can obtain 
%
\begin{eqnarray}
  \myZero1{k}{\ell} & := & 
  \mydet{A}{k}{\ell-1}   \, \mydet{F}{k}{\ell+1}(\zeta) 
  - \mydet{A}{k}{\ell}   \, \mydet{F}{k}{\ell}(\zeta) 
  + \zeta\mydet{A}{k-1}{\ell} \, \mydet{F}{k+1}{\ell}(\zeta) 
  = 0 
\\[2mm]
  \myZero3{k}{\ell} & := & 
    \mydet{A}{k}{\ell-1}    \, \mydet{F}{k+1}{\ell+1}(\zeta) 
  - \mydet{A}{k+1}{\ell}    \, \mydet{F}{k}{\ell}(\zeta) 
  + \zeta\mydet{A}{k}{\ell} \, \mydet{F}{k+1}{\ell}(\zeta) 
  = 0 
\\[2mm]
  \myZero4{k}{\ell} & := & 
    \mydet{F}{k+1}{\ell+1}(\zeta) \,\partial_{+} \mydet{A}{k}{\ell-1} 
  - \mydet{A}{k+1}{\ell}          \,\partial_{+} \mydet{F}{k}{\ell}(\zeta) 
  + \zeta \mydet{A}{k}{\ell}      \,\partial_{+} \mydet{F}{k+1}{\ell}(\zeta) 
  = 0 
\\[2mm]
  \myZero6{k}{\ell} & := & 
  \mydet{F}{k}{\ell+1}(\zeta)  \,\partial_{-} \mydet{A}{k}{\ell-1} 
  - \mydet{A}{k}{\ell}         \,\partial_{-} \mydet{F}{k}{\ell}(\zeta) 
  + \zeta \mydet{A}{k-1}{\ell} \,\partial_{-} \mydet{F}{k+1}{\ell}(\zeta) 
  = 0 
\end{eqnarray}
%
correspondingly. These identities, combined with 
\begin{equation}
  \Delta^{k}_{\ell}:= 
  \left( \mydet{A}{k}{\ell} \right)^{2} 
  - \mydet{A}{k}{\ell-1} \mydet{A}{k}{\ell+1} 
  - \mydet{A}{k-1}{\ell} \mydet{A}{k+1}{\ell} 
  = 0
\end{equation}
lead to 
%
\begin{eqnarray}
  \myZero0{k}{\ell} & := &
  \mydet{A}{k}{\ell} \mydet{F}{k}{\ell+1} 
  - \mydet{A}{k}{\ell+1} \mydet{F}{k}{\ell} 
  - \mydet{A}{k-1}{\ell} \mydet{F}{k+1}{\ell+1} 
  = 0 
\\[2mm]
  \myZero2{k}{\ell} & := & 
  \mydet{A}{k}{\ell} \mydet{F}{k+1}{\ell+1} 
  - \mydet{A}{k+1}{\ell} \mydet{F}{k}{\ell+1} 
  + \zeta\mydet{A}{k}{\ell+1} \mydet{F}{k+1}{\ell} 
  = 0. 
\end{eqnarray}
%
Expanding the above equations, 
%
\begin{eqnarray}
  \myZero0{k}{\ell} & = & 
  \Delta^{k}_{l} 
  + \zeta\,\mybarzero0{k}{\ell} 
  + ... + 
  (-\zeta)^{\ell-1}\myzero0{k}{\ell} 
\\
  \myZero1{k}{\ell} & = &
  \zeta\,\mybarzero1{k}{\ell} 
  + ... + 
  (-\zeta)^{\ell-1}\myzero1{k}{\ell} 
\\
  \myZero2{k}{\ell} & = &
  \zeta\,\mybarzero2{k}{\ell} 
  + ... + 
  (-\zeta)^{\ell}\Delta^{k}_{\ell} 
\\
  \myZero3{k}{\ell} & = &
  \zeta\,\mybarzero3{k}{\ell} 
  + ... + 
  (-\zeta)^{\ell-1}\myzero3{k}{\ell}, 
\end{eqnarray}
%
where 
%
\begin{eqnarray}
  \myzero0{k}{\ell} & = &
    D_{+}\, \mydet{A}{k-1}{\ell} \cdot \mydet{A}{k}{\ell} 
    - \mydet{A}{k-1}{\ell-1} \mydet{A}{k}{\ell+1} 
\\
  \myzero1{k}{\ell} & = &
    D_{+}\, \mydet{A}{k-1}{\ell} \cdot \mydet{A}{k}{\ell-1} 
    - \mydet{A}{k-1}{\ell-1} \mydet{A}{k}{\ell} 
\\
  \myzero3{k}{\ell} & = &
    D_{+}\, \mydet{A}{k}{\ell} \cdot \mydet{A}{k}{\ell-1} 
    - \mydet{A}{k-1}{\ell-1} \mydet{A}{k+1}{\ell} 
\end{eqnarray}
%
with 
%
\begin{eqnarray}
  \mybarzero0{k}{\ell} & = &
    - \mydet{A}{k}{\ell} \partial_{-}\mydet{A}{k}{\ell} 
    + \mydet{A}{k}{\ell+1} \partial_{-}\mydet{A}{k}{\ell-1} 
    + \mydet{A}{k-1}{\ell} \partial_{-}\mydet{A}{k+1}{\ell} 
\\
  \mybarzero1{k}{\ell} & = &
    D_{-}\, \mydet{A}{k}{\ell-1} \cdot \mydet{A}{k}{\ell} 
    + \mydet{A}{k-1}{\ell} \mydet{A}{k+1}{\ell-1} 
\\
  \mybarzero2{k}{\ell} & = &
    D_{-}\, \mydet{A}{k}{\ell} \cdot \mydet{A}{k+1}{\ell} 
    + \mydet{A}{k+1}{\ell-1} \mydet{A}{k}{\ell+1} 
\\
  \mybarzero3{k}{\ell} & = &
    D_{-}\, \mydet{A}{k}{\ell-1} \cdot \mydet{A}{k+1}{\ell} 
    + \mydet{A}{k+1}{\ell-1} \mydet{A}{k}{\ell}, 
\end{eqnarray}
%
and introducing 
%
\begin{eqnarray}
  \myZero5{k}{\ell} 
  & := & 
  \partial_{+}\myZero3{k}{\ell} - \myZero4{k}{\ell} = 0
\\
  \myZero7{k}{\ell} 
  & := & 
  \partial_{-}\myZero1{k}{\ell} - \myZero6{k}{\ell} = 0
\end{eqnarray}
%
one can derive by straightforward algebra the following identities:
%
\begin{eqnarray}
\hspace{-10mm}
  \mydet{A}{k}{\ell-1} 
  \left( 
    D_{+}\, \mydet{F}{k+1}{\ell+1} \cdot \mydet{A}{k}{\ell} 
    + \mydet{A}{k+1}{\ell+1} \mydet{F}{k}{\ell} 
  \right) 
  & = & 
    \mydet{A}{k}{\ell} \myZero5{k}{\ell} 
  - \left( \partial_{+}\mydet{A}{k}{\ell} \right) \myZero3{k}{\ell} 
  - \mydet{F}{k}{\ell} \myzero0{k+1}{\ell} 
\\[2mm] 
\hspace{-10mm}
  \mydet{A}{k-1}{\ell-1} 
  \left( 
    D_{+}\, \mydet{F}{k}{\ell+1} \cdot \mydet{A}{k}{\ell} 
    + \zeta\mydet{A}{k}{\ell+1} \mydet{F}{k}{\ell} 
  \right) 
  & = & 
    \mydet{A}{k}{\ell} \myZero5{k-1}{\ell} 
  - \left( \partial_{+}\mydet{A}{k}{\ell} \right) \myZero3{k-1}{\ell} 
  - \zeta\mydet{F}{k}{\ell} \myzero0{k}{\ell} 
\\[2mm]
\hspace{-10mm}
  \mydet{A}{k}{\ell-2} 
  \left( 
    D_{+}\, \mydet{F}{k+1}{\ell} \cdot \mydet{A}{k}{\ell} 
    + \mydet{A}{k+1}{\ell} \mydet{F}{k}{\ell} 
  \right) 
  & = & 
    \mydet{A}{k}{\ell} \myZero5{k}{\ell-1} 
  - \left( \partial_{+}\mydet{A}{k}{\ell} \right) \myZero3{k}{\ell-1} 
  + \mydet{A}{k+1}{\ell} \myZero1{k}{\ell-1} 
\nonumber\\&& 
  - \mydet{F}{k}{\ell-1} \myzero1{k+1}{\ell} 
  + \zeta\mydet{F}{k+1}{\ell-1} \myzero3{k}{\ell} 
\end{eqnarray}
%
and 
%
\begin{eqnarray}
\hspace{-10mm}
  \mydet{A}{k+1}{\ell-1} 
  \left( 
    D_{-}\, \mydet{F}{k+1}{\ell+1} \cdot \mydet{A}{k}{\ell} 
    - \mydet{A}{k}{\ell+1} \mydet{F}{k+1}{\ell} 
  \right) 
  & = & 
    \mydet{A}{k}{\ell} \myZero7{k+1}{\ell} 
  - \left( \partial_{-}\mydet{A}{k}{\ell} \right) \myZero1{k+1}{\ell} 
  - \mydet{F}{k+1}{\ell} \mybarzero2{k}{\ell} 
\\[2mm] 
\hspace{-10mm}
  \mydet{A}{k}{\ell-1} 
  \left( 
    D_{-}\, \mydet{F}{k}{\ell+1} \cdot \mydet{A}{k}{\ell} 
    - \zeta\mydet{A}{k-1}{\ell+1} \mydet{F}{k+1}{\ell} 
  \right) 
  & = & 
    \mydet{A}{k}{\ell} \myZero7{k}{\ell} 
  - \left( \partial_{-}\mydet{A}{k}{\ell} \right) \myZero1{k}{\ell} 
  - \zeta\mydet{F}{k+1}{\ell} \mybarzero2{k-1}{\ell} 
\\[2mm] 
\hspace{-10mm}
  \mydet{A}{k}{\ell-2} 
  \left( 
    D_{-}\, \mydet{F}{k}{\ell} \cdot \mydet{A}{k}{\ell} 
    + \mydet{A}{k-1}{\ell} \mydet{F}{k+1}{\ell} 
  \right) 
  & = & 
    \mydet{A}{k}{\ell} \myZero7{k}{\ell-1} 
  + \mydet{A}{k-1}{\ell} \myZero3{k}{\ell-1} 
  - \left( \partial_{-}\mydet{A}{k}{\ell} \right) \myZero1{k}{\ell-1} 
\nonumber\\&& 
  + \mydet{F}{k}{\ell-1} \mybarzero1{k}{\ell} 
  - \zeta\mydet{F}{k+1}{\ell-1} \mybarzero3{k-1}{\ell} 
\end{eqnarray}
%
which means 
%
\begin{eqnarray}
\label{app-pos-a}
    D_{+}\, \mydet{F}{k+1}{\ell+1} \cdot \mydet{A}{k}{\ell} 
    + \mydet{A}{k+1}{\ell+1} \mydet{F}{k}{\ell} 
  & = & 0
\\
    D_{+}\, \mydet{F}{k}{\ell+1} \cdot \mydet{A}{k}{\ell} 
    + \zeta\mydet{A}{k}{\ell+1} \mydet{F}{k}{\ell} 
  & = & 0 
\\
    D_{+}\, \mydet{F}{k+1}{\ell} \cdot \mydet{A}{k}{\ell} 
    + \mydet{A}{k+1}{\ell} \mydet{F}{k}{\ell} 
  & = & 0 
\end{eqnarray}
%
and 
%
\begin{eqnarray}
    D_{-}\, \mydet{F}{k+1}{\ell+1} \cdot \mydet{A}{k}{\ell} 
    - \mydet{A}{k}{\ell+1} \mydet{F}{k+1}{\ell} 
  & = & 0
\\[2mm] 
    D_{-}\, \mydet{F}{k}{\ell+1} \cdot \mydet{A}{k}{\ell} 
    - \zeta\mydet{A}{k-1}{\ell+1} \mydet{F}{k+1}{\ell} 
  & = & 0 
\\[2mm] 
    D_{-}\, \mydet{F}{k}{\ell} \cdot \mydet{A}{k}{\ell} 
    + \mydet{A}{k-1}{\ell} \mydet{F}{k+1}{\ell} 
  & = & 0. 
\label{app-neg-c}
\end{eqnarray}
%
These equations, 
when rewritten in terms of $\mydet{B}{k}{\ell}$ given by \eref{def-B}, 
are nothing but \eref{AB-pos-a}--\eref{AB-neg-c}, while in terms of  
$\mydet{C}{k}{\ell}$ given by \eref{def-C} they become 
%
\begin{eqnarray}
  0 & = & 
  D_{+} \, \mydet{C}{k}{\ell} \cdot \mydet{A}{k}{\ell} 
  + \zeta \mydet{C}{k}{\ell-1} \, \mydet{A}{k}{\ell+1} 
\\[2mm] 
  0 & = & 
  D_{+} \, \mydet{C}{k+1}{\ell} \cdot \mydet{A}{k}{\ell} 
  + \mydet{A}{k+1}{\ell+1} \, \mydet{C}{k}{\ell-1} 
\\[2mm] 
  0 & = & 
  D_{+} \, \mydet{C}{k+1}{\ell} \cdot \mydet{A}{k}{\ell+1} 
  + \mydet{A}{k+1}{\ell+1} \, \mydet{C}{k}{\ell} 
\end{eqnarray}
%
and 
%
\begin{eqnarray}
  0 & = & 
  D_{-} \, \mydet{C}{k}{\ell} \cdot \mydet{A}{k}{\ell} 
  - \zeta \mydet{A}{k-1}{\ell+1} \, \mydet{C}{k+1}{\ell-1} 
\\[2mm] 
  0 & = & 
  D_{-} \, \mydet{C}{k}{\ell} \cdot \mydet{A}{k}{\ell+1} 
  + \mydet{A}{k-1}{\ell+1} \, \mydet{C}{k+1}{\ell} 
\\[2mm] 
  0 & = & 
  D_{-} \, \mydet{C}{k+1}{\ell} \cdot \mydet{A}{k}{\ell} 
  - \mydet{A}{k}{\ell+1} \, \mydet{C}{k+1}{\ell-1}. 
\end{eqnarray}
%
The above formulae are enough to obtain the Toeplitz solutions presented in 
section \ref{sec-toeplitz}.

\section*{References}

\end{document}